\documentclass[11pt]{article}
\usepackage{microtype}
\usepackage{amsmath}
\usepackage{amssymb}
\usepackage{bm}
\usepackage{graphics}

\newcommand{\be}{\begin{equation}}
\newcommand{\ee}{\end{equation}}
\newcommand{\om}{\omega}
\newcommand{\vep}{\varepsilon}
\newcommand{\ra}{\rightarrow}
\newcommand{\D}{\mathrm{d}}

\newcommand{\E}{\mathrm{e}}
\newcommand{\reals}{\mathbb{R}}
\newcommand{\cX}{\mathcal{X}}
\newcommand{\bk}{\mathbf{k}}

\newcommand{\bL}{\mathbf{L}}
\newcommand{\bmu}{\bm\mu}

\newcommand{\bx}{\mathbf{x}}
\newcommand{\hp}{\hat p}
\newcommand{\hq}{\hat q}

\newcommand{\hlambda}{\hat\lambda}

\newcommand{\var}{\text{var}}
\newcommand{\err}{\text{err}}
\newcommand{\bom}{\bm\om}
\newcommand{\tPi}{\tilde\Pi}
\newcommand{\tpi}{\tilde\pi}
\newcommand{\tG}{\tilde G}
\newcommand{\id}{\mathbf{1}}
\newcommand{\emp}{\textbf}
\newcommand{\exdiff}[1]{$\!\!^{\mathbf{[#1]}}$}

\numberwithin{equation}{section}
\numberwithin{figure}{section}

\newcounter{exer}
\numberwithin{exer}{subsection}
\newenvironment{exercise}
{
 \begin{list}{\textbf{\thesubsection.\arabic{exer}.}~}
 {
  \usecounter{exer}
  \setlength{\labelsep}{0pt}
  \setlength{\itemindent}{0pt}
  \setlength{\labelwidth}{0pt}
  \setlength{\leftmargin}{0pt}
 }
}
{\end{list}}

\begin{document}

\title{\vspace*{-1.2in}
\textbf{A basic introduction to large deviations: Theory, applications, simulations}\footnote{Published in R. Leidl and A. K. Hartmann (eds), \textit{Modern Computational Science 11: Lecture Notes from the 3rd International Oldenburg Summer School}, BIS-Verlag der Carl von Ossietzky Universit\"at Oldenburg, 2011.}
}

\author{
Hugo Touchette\\
\textit{School of Mathematical Sciences, Queen Mary, University of London}\\
\textit{London E1 4NS, United Kingdom}
}

\date{\today}
\maketitle

\begin{abstract}
The theory of large deviations deals with the probabilities of rare events (or fluctuations) that are exponentially small as a function of some parameter, e.g., the number of random components of a system, the time over which a stochastic system is observed, the amplitude of the noise perturbing a dynamical system or the temperature of a chemical reaction. The theory has applications in many different scientific fields, ranging from queuing theory to statistics and from finance to engineering. It is also increasingly used in statistical physics for studying both equilibrium and nonequilibrium systems. In this context, deep analogies can be made between familiar concepts of statistical physics, such as the entropy and the free energy, and concepts of large deviation theory having more technical names, such as the rate function and the scaled cumulant generating function.

The first part of these notes introduces the basic elements of large deviation theory at a level appropriate for advanced undergraduate and  graduate students in physics, engineering, chemistry, and mathematics. The focus there is on the simple but powerful ideas behind large deviation theory, stated in non-technical terms, and on the application of these ideas in simple stochastic processes, such as sums of independent and identically distributed random variables and Markov processes. Some physical applications of these processes are covered in exercises contained at the end of each section. 

In the second part, the problem of numerically evaluating large deviation probabilities is treated at a basic level. The fundamental idea of importance sampling is introduced there together with its sister idea, the exponential change of measure. Other numerical methods based on sample means and generating functions, with applications to Markov processes, are also covered. 
\end{abstract}
\vfill

\newpage
\tableofcontents

\newpage
\section{Introduction}
\label{secintro}

The goal of these lecture notes, as the title says, is to give a basic introduction to the theory of large deviations at three levels: theory, applications and simulations. The notes follow closely my recent review paper on large deviations and their applications in statistical mechanics \cite{touchette2009}, but are, in a way, both narrower and wider in scope than this review paper.

They are narrower, in the sense that the mathematical notations have been cut down to a minimum in order for the theory to be understood by advanced undergraduate and graduate students in science and engineering, having a basic background in probability theory and stochastic processes (see, e.g., \cite{grimmett2001}). The simplification of the mathematics amounts essentially to two things: i) to focus on random variables taking values in $\reals$ or $\reals^D$, and ii) to state all the results in terms of probability densities, and so to assume that probability densities always exist, if only in a weak sense. These simplifications are justified for most applications and are convenient for conveying the essential ideas of the theory in a clear way, without the hindrance of technical notations.

These notes also go beyond the review paper \cite{touchette2009}, in that they cover subjects not contained in that paper, in particular the subject of numerical estimation or \textit{simulation} of large deviation probabilities. This is an important subject that I intend to cover in more depth in the future.

Sections~\ref{secnum1} and \ref{secnum2} of these notes are a first and somewhat brief attempt in this direction. Far from covering all the methods that have been developed to simulate large deviations and rare events, they concentrate on the central idea of large deviation simulations, namely that of exponential change of measure, and they elaborate from there on certain simulation techniques that are easily applicable to sums of independent random variables, Markov chains, stochastic differential equations and continuous-time Markov processes in general.

Many of these applications are covered in the exercises contained at the end of each section. The level of difficulty of these exercises is quite varied: some are there to practice the material presented, while others go beyond that material and may take hours, if not days or weeks, to solve completely. For convenience, I have rated them according to Knuth's logarithmic scale.\footnote{00 = Immediate; 10 = Simple; 20 = Medium; 30 = Moderately hard; 40 = Term project; 50 = Research problem. See the superscript attached to each exercise.}

In closing this introduction, let me emphasize again that these notes are not meant to be complete in any way. For one thing, they lack the rigorous notation needed for handling large deviations in a precise mathematical way, and only give a hint of the vast subject that is large deviation and rare event simulation. On the simulation side, they also skip over the subject of error estimates, which would fill an entire section in itself. In spite of this, I hope that the material covered will have the effect of enticing readers to learn more about large deviations and of conveying a sense of the wide applicability, depth and beauty of this subject, both at the theoretical and computational levels.

\section{Basic elements of large deviation theory}
\label{sectheory}

\subsection{Examples of large deviations}

We  start our study of large deviation theory by considering a sum of real random variables (RV for short) having the form
\be
S_n=\frac{1}{n}\sum_{i=1}^n X_i.
\ee
Such a sum is often referred to in mathematics or statistics as a \emp{sample mean}. We are interested in computing the \emp{probability density function} $p_{S_n}(s)$ (pdf for short) of $S_n$ in the simple case where the $n$ RVs are mutually \emp{independent and identically distributed} (IID for short).\footnote{The pdf of $S_n$ will be denoted by $p_{S_n}(s)$ or more simply by $p(S_n)$ when no confusion arises.} This means that the joint pdf of $X_1,\ldots,X_n$ factorizes as follows:
\be
p(X_1,\ldots,X_n)=\prod_{i=1}^n p(X_i),
\ee
with $p(X_i)$ a fixed pdf for each of the $X_i$'s.\footnote{To avoid confusion, we should write $p_{X_i}(x)$, but in order not to overload the notation, we will simply write $p(X_i)$. The context should make clear what RV we are referring to.} 

We compare next two cases for $p(X_i)$:
\begin{itemize}
\item \emp{Gaussian pdf:}
\be
p(x)=\frac{1}{\sqrt{2\pi\sigma^2}} \E^{-(x-\mu)^2/(2\sigma^2)},\quad x\in\reals,
\ee
where $\mu=E[X]$ is the \emp{mean} of $X$ and $\sigma^2=E[(X-\mu)^2]$ its \emp{variance}.\footnote{The symbol $E[X]$ denotes the expectation or expected value of $X$, which is often denoted by $\langle X\rangle$ in physics.}

\item \emp{Exponential pdf:}
\be
p(x)=\frac{1}{\mu} \E^{-x/\mu},\quad x\in [0,\infty),
\label{eqexpd1}
\ee
with mean $E[X]=\mu>0$. 
\end{itemize}

What is the form of $p_{S_n}(s)$ for each pdf?

To find out, we write the pdf associated with the event $S_n=s$ by summing the pdf of all the values or \emp{realizations} $(x_1,\ldots,x_n)\in\reals^n$ of $X_1,\ldots,X_n$ such that $S_n=s$.\footnote{Whenever possible, random variables will be denoted by uppercase letters, while their values or realizations will be denoted by lowercase letters. This follows the convention used in probability theory. Thus, we will write $X=x$ to mean that the RV $X$ takes the value $x$.} In terms of Dirac's delta function $\delta(x)$, this is written as
\be
p_{S_n}(s)=\int_\reals \D x_1\cdots \int_\reals \D x_n\ \delta\big(\textstyle\sum_{i=1}^n x_i -ns\big)\ p(x_1,\ldots,x_n).
\label{eqrep1}
\ee
From this expression, we can then obtain an explicit expression for $p_{S_n}(s)$ by using the method of generating functions (see Exercise~\ref{excf1}) or by substituting the Fourier representation of $\delta(x)$ above and by explicitly evaluating the $n+1$ resulting integrals. The result obtained for both the Gaussian and the exponential densities has the general form
\be
p_{S_n}(s)\approx \E^{-nI(s)},
\label{eqldt1}
\ee
where 
\be
I(s)=\frac{(x-\mu)^2}{2\sigma^2},\qquad s\in\reals
\label{eqrfg1}
\ee
for the Gaussian pdf, whereas
\be
I(s)=\frac{s}{\mu}-1-\ln\frac{s}{\mu},\qquad s\geq 0
\label{eqldexp1}
\ee
for the exponential pdf.

\begin{figure}[t]
\resizebox{\textwidth}{!}{\includegraphics{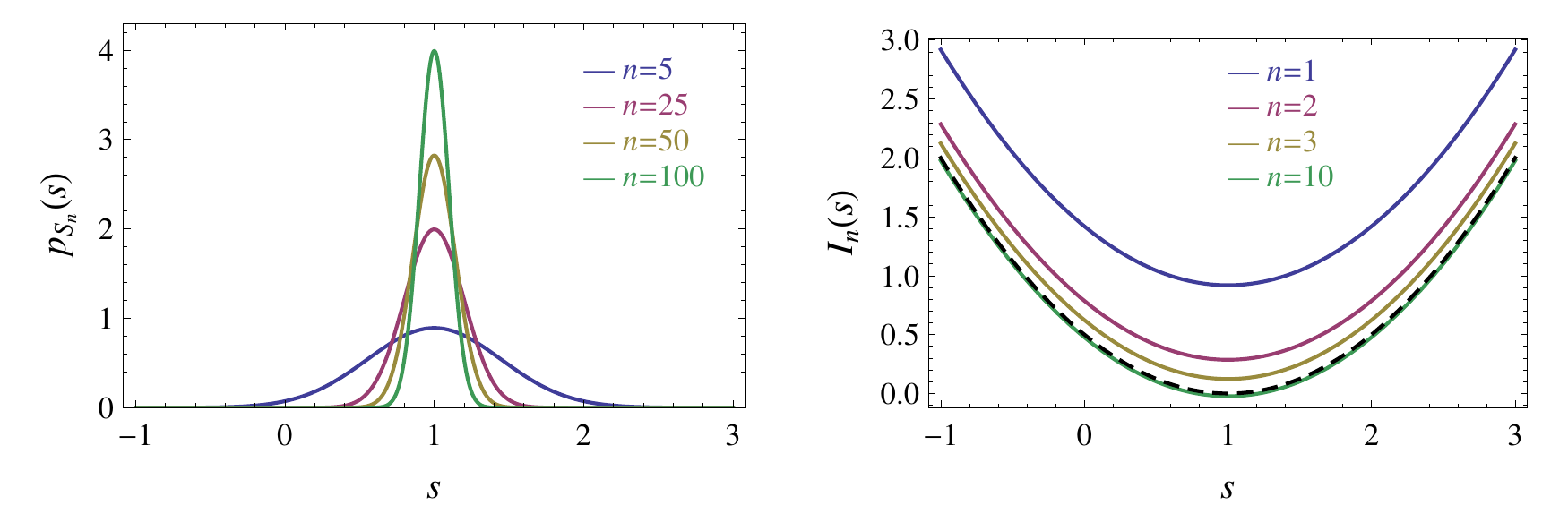}}
\caption{(Left) pdf $p_{S_n}(s)$ of the Gaussian sample mean $S_n$ for $\mu=\sigma=1$. (Right) $I_n(s)=-\frac{1}{n}\ln p_{S_n}(s)$ for different values of $n$ demonstrating a rapid convergence towards the rate function $I(s)$ (dashed line).}
\label{figgaussianld1}
\end{figure}

We will come to understand the meaning of the approximation sign ($\approx$) more clearly in the next subsection. For now we just take it as meaning that the dominant behaviour of $p(S_n)$ as a function of $n$ is a decaying exponential in $n$. Other terms in $n$ that may appear in the exact expression of $p(S_n)$ are sub-exponential in $n$.

The picture of the behaviour of $p(S_n)$ that we obtain from the result of (\ref{eqldt1}) is that $p(S_n)$ decays to $0$ exponentially fast with $n$ for all values $S_n=s$ for which the function $I(s)$, which controls the rate of decay of $p(S_n)$, is positive. But notice that $I(s)\geq 0$ for both the Gaussian and exponential densities, and that $I(s)=0$ in both cases only for $s=\mu=E[X_i]$. Therefore, since the pdf of $S_n$ is normalized, it must get more and more concentrated around the value $s=\mu$ as $n\ra\infty$ (see Fig.~\ref{figgaussianld1}), which means that $p_{S_n}(s)\ra \delta(s-\mu)$ in this limit. We say in this case that $S_n$ converges in probability or in density to its mean. 

As a variation of the Gaussian and exponential samples means, let us now consider a sum of \emp{discrete random variables} having a probability distribution $P(X_i)$ instead of continuous random variables with a pdf $p(X_i)$.\footnote{Following the convention in probability theory, we use the lowercase $p$ for continuous probability densities and the uppercase $P$ for discrete probability distributions and probability assignments in general. Moreover, following the notation used before, we will denote the distribution of a discrete $S_n$ by $P_{S_n}(s)$ or simply $P(S_n)$.} To be specific, consider the case of IID \emp{Bernoulli RVs} $X_1,\ldots,X_n$ taking values in the set $\{0,1\}$ with probability $P(X_i=0)=1-\alpha$ and $P(X_i=1)=\alpha$. What is the form of the probability distribution $P(S_n)$ of $S_n$ in this case?

Notice that we are speaking of a probability distribution because $S_n$ is now a discrete variable taking values in the set $\{0,1/n,2/n,\ldots,(n-1)/n,1\}$. In the previous Gaussian and exponential examples, $S_n$ was a continuous variable characterized by its pdf $p(S_n)$. 

With this in mind, we can obtain the exact expression of $P(S_n)$ using methods similar to those used to obtain $p(S_n)$ (see Exercise~\ref{exbs1}). The  result  is different from that found for the Gaussian and exponential densities, but what is remarkable is that the distribution of the Bernoulli sample mean also contains a dominant exponential term having the form
\be
P_{S_n}(s)\approx \E^{-nI (s)},
\label{eqldbern1}
\ee 
where $I(s)$ is now given by
\be
I(s)=s\ln \frac{s}{\alpha} +(1-s)\ln \frac{1-s}{1-\alpha},\qquad s\in [0,1].
\label{eqbern1}
\ee

\begin{figure}[t]
\resizebox{\textwidth}{!}{\includegraphics{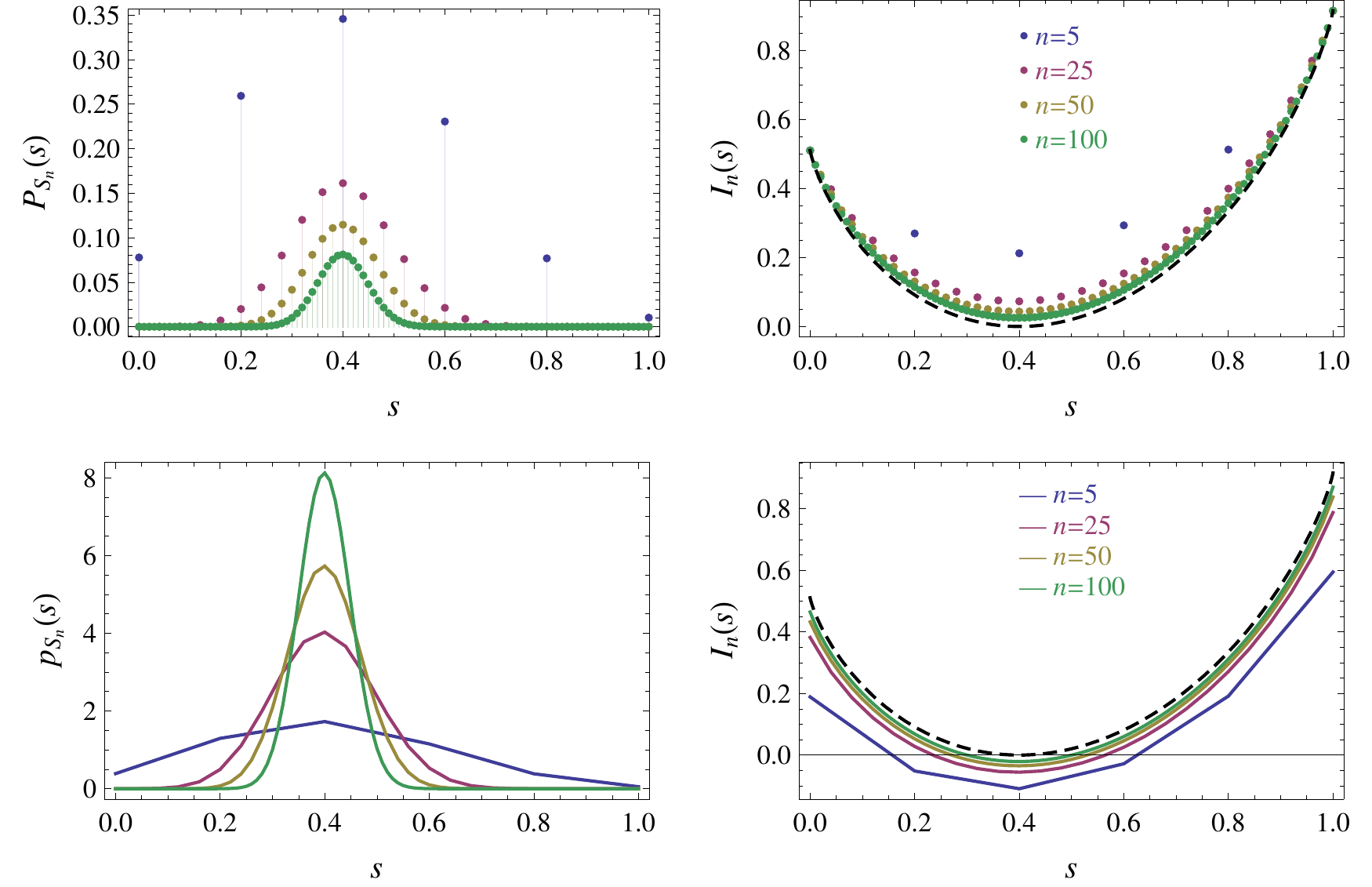}}
\caption{(Top left) Discrete probability distribution $P_{S_n}(s)$ of the Bernoulli sample mean for $\alpha=0.4$ and different values of $n$. (Top right) Finite-$n$ rate function $I_n(s)=-\frac{1}{n}\ln P_{S_n}(s)$. The rate function $I(s)$ is the dashed line. (Bottom left). Coarse-grained pdf $p_{S_n}(s)$ for the Bernoulli sample mean. (Bottom right) $I_n(s)=-\frac{1}{n}\ln p_{S_n}(s)$ as obtained from the coarse-grained pdf.}
\label{figbernld1}
\end{figure}

The behaviour of the exact expression of $P(S_n)$ as $n$ grows is shown in Fig.~\ref{figbernld1} together with the plot of $I(s)$ as given by Eq.~(\ref{eqbern1}). Notice how $P(S_n)$ concentrates around its mean $\mu=E[X_i]=\alpha$ as a result of the fact that $I(s)\geq 0$ and that $s=\mu$ is the only value of $S_n$ for which $I=0$. Notice also how the support of $P(S_n)$ becomes ``denser'' as $n\ra\infty$, and compare this property with the fact that $I(s)$ is a continuous function despite $S_n$ being discrete. Should $I(s)$ not be defined for discrete values if $S_n$ is a discrete RV? We address this question next.

\subsection{The large deviation principle}

The general exponential form $\E^{-nI(s)}$ that we found for our three previous sample means (Gaussian, exponential and Bernoulli) is the founding result or property of large deviation theory, referred to as the large deviation principle. The reason why a whole theory can be built on such a seemingly simple result is that it arises in the context of many stochastic processes, not just IID sample means, as we will come to see in the next sections, and as can be seen from other contributions to this volume; see, e.g., Engel's.\footnote{The volume referred to is the volume of lecture notes produced for the summer school; see \texttt{http://www.mcs.uni-oldenburg.de/}}

The rigorous, mathematical definition of the large deviation principle involves many concepts of topology and measure theory that are too technical to be presented here (see Sec.~3.1 and Appendix B of \cite{touchette2009}). For simplicity, we will say here that a random variable $S_n$ or its pdf $p(S_n)$ satisfies a \emp{large deviation principle} (LDP) if the following limit exists:
\be
\lim_{n\ra\infty} -\frac{1}{n}\ln p_{S_n}(s)=I(s)
\label{eqldlim1}
\ee
and gives rise to a function $I(s)$, called the \emp{rate function}, which is not everywhere zero.

The relation between this definition and the approximation notation used earlier should be clear: the fact that the behaviour of $p_{S_n}(s)$ is dominated for large $n$ by a decaying exponential means that the exact pdf of $S_n$ can be written as
\be
p_{S_n}(s)=\E^{-nI(s)+o(n)}
\ee
where $o(n)$ stands for any correction term that is sub-linear in $n$. Taking the large deviation limit of (\ref{eqldlim1}) then yields
\be
\lim_{n\ra\infty} -\frac{1}{n}\ln p_{S_n}(s)=I(s)-\lim_{n\ra\infty} \frac{o(n)}{n} = I(s),
\ee
since $o(n)/n\ra 0$. We see therefore that the large deviation limit of Eq.~(\ref{eqldlim1}) is the limit needed to retain the dominant exponential term in $p(S_n)$ while discarding any other sub-exponential terms. For this reason, large deviation theory is often said to be concerned with estimates of probabilities on the logarithmic scale. 

This point is illustrated in Fig.~\ref{figbernld1}. There we see that the function 
\be
I_n(s)=-\frac{1}{n}\ln p_{S_n}(s)
\ee 
is not quite equal to the limiting rate function $I(s)$, because of terms of order $o(n)/n=o(1)$; however, it does converge to $I(s)$ as $n$ gets larger. Plotting $p(S_n)$ on this scale therefore reveals the rate function in the limit of large $n$. This convergence will be encountered repeatedly in the sections on the numerical evaluation of large deviation probabilities.

It should be emphasized again that the definition of the LDP given above is a simplification of the rigorous definition used in mathematics, due to the mathematician Srinivas Varadhan.\footnote{Recipient of the 2007 Abel Prize for his ``fundamental contributions to probability theory and in particular for creating a unified theory of large deviations.''} The real, mathematical definition is expressed in terms of probability measures of certain sets rather than in terms of probability densities, and involves upper and lower bounds on these probabilities rather than a simple limit (see Sec.~3.1 and Appendix B of \cite{touchette2009}). The mathematical definition also applies a priori to any RVs, not just continuous RVs with a pdf.

In these notes, we will simplify the mathematics by assuming that the random variables or stochastic processes that we study have a pdf. In fact, we will often assume that pdfs exist even for RVs that are not continuous but ``look'' continuous at some scale.

To illustrate this point, consider again the Bernoulli sample mean. We noticed that $S_n$ in this case is not a continuous RV, and so does not have a pdf. However, we also noticed that the values that $S_n$ can take become dense in the interval $[0,1]$ as $n\ra\infty$, which means that the support of the discrete probability distribution $P(S_n)$ becomes dense in this limit, as shown in Fig.~\ref{figbernld1}. From a practical point of view, it therefore makes sense to treat $S_n$ as a continuous RV for large $n$ by interpolating $P(S_n)$ to a continuous function $p(S_n)$ representing the ``probability density'' of $S_n$.

This pdf is obtained in general simply by considering the probability $P(S_n\in [s,s+\Delta s])$ that $S_n$ takes a value in a tiny interval surrounding the value $s$, and by then dividing this probability by the ``size'' $\Delta s$ of that interval:
\be
p_{S_n}(s)=\frac{P(S_n\in [s,s+\Delta s])}{\Delta s}.
\ee
The pdf obtained in this way is referred to as a \emp{smoothed}, \emp{coarse-grained} or \emp{weak density}.\footnote{\label{fnldp1}To be more precise, we should make clear that the coarse-grained pdf depends on the spacing $\Delta s$ by writing, say, $p_{S_n,\Delta s}(s)$. However, to keep the notation to a minimum, we will use the same lowercase $p$ to denote a coarse-grained pdf and a ``true'' continuous pdf. The context should make clear which of the two is used.} In the case of the Bernoulli sample mean, it is simply given by
\be
p_{S_n}(s)=nP_{S_n}(s)
\ee
if the spacing $\Delta s$ between the values of $S_n$ is chosen to be $1/n$.

This process of replacing a discrete variable by a continuous one in some limit or at some scale is known in physics as the \emp{continuous} or \emp{continuum limit} or as the \emp{macroscopic limit}. In mathematics, this limit is expressed  via the notion of weak convergence (see \cite{dupuis1997}).

\subsection{The G\"artner-Ellis Theorem}

Our goal in the next sections will be to study random variables and stochastic processes that satisfy an LDP and to find analytical as well as numerical ways to obtain the corresponding rate function. There are many ways whereby a random variable, say $S_n$, can be shown to satisfy an LDP:
\begin{itemize}
\item \emp{Direct method:} Find the expression of $p(S_n)$ and show that it has the form of the LDP;
\item \emp{Indirect method:} Calculate certain functions of $S_n$, such as generating functions, whose properties can be used to infer that $S_n$ satisfies an LDP;
\item \emp{Contraction method:} Relate $S_n$ to another random variable, say $A_n$, which is known to satisfy an LDP and derive from this an LDP for $S_n$.
\end{itemize}

We have used the first method when discussing the Gaussian, exponential and Bernoulli sample means. 

The main result of large deviation theory that we will use in these notes to obtain LDPs along the indirect method is called the \emp{G\"artner-Ellis Theorem} (GE Theorem for short), and is based on the calculation of the following function:
\be
\lambda(k)=\lim_{n\ra\infty}\frac{1}{n}\ln E[\E^{nk S_n}],
\label{eqscgf1}
\ee
known as the \emp{scaled cumulant generating function}\footnote{The function $E[\E^{kX}]$ for $k$ real is known as the \emp{generating function} of the RV $X$; $\ln E[\E^{kX}]$ is known as the \emp{log-generating function} or \emp{cumulant generating function}. The word ``scaled'' comes from the extra factor $1/n$.} (SCGF for short). In this expression $E[\cdot]$ denotes the expected value, $k$ is a real parameter, and $S_n$ is an arbitrary RV; it is not necessarily an IID sample mean or even a sample mean. 

The point of the GE Theorem is to be able to calculate $\lambda(k)$ without knowing $p(S_n)$. We will see later that this is possible. Given $\lambda(k)$, the GE Theorem then says that, if $\lambda(k)$ is differentiable,\footnote{The statement of the GE Theorem given here is a simplified version of the full result, which is essentially the result proved by G\"artner \cite{gartner1977}. See Sec.~3.3 of \cite{touchette2009} for a more complete presentation of GE Theorem and \cite{ellis1984} for a rigorous account of it.} then 
\begin{itemize}
\item $S_n$ satisfies an LDP, i.e.,
\be
\lim_{n\ra\infty}-\frac{1}{n}\ln p_{S_n}(s)=I(s);
\ee
\item The rate function $I(s)$ is given by the \emp{Legendre-Fenchel transform} of $\lambda(k)$:
\be
I(s)=\sup_{k\in\reals}\{ks-\lambda(k)\},
\label{eqlf1}
\ee
where ``$\sup$'' stands for the \emp{supremum}.\footnote{In these notes, a $\sup$ can be taken to mean the same as a $\max$.}
\end{itemize}

We will come to calculate $\lambda(k)$ and its Legendre-Fenchel transform for specific RVs in the next section. It will also be seen in there that $\lambda(k)$ does not always exist (this typically happens when the pdf of a random variable does not admit an LDP). Some exercises at the end of this section illustrate many useful properties of $\lambda(k)$ when this function exists and is twice differentiable.

\subsection{Varadhan's Theorem}

The rigorous proof of the GE Theorem is too technical to be presented here. However, there is a way to justify this result by deriving in a  heuristic way another result known as Varadhan's Theorem.

The latter theorem is concerned with the evaluation of a functional\footnote{A function of a function is called a \emp{functional}.} expectation of the form
\be
W_n[f]=E[\E^{nf(S_n)}]=\int_\reals p_{S_n}(s)\, \E^{nf(s)}\, \D s,
\ee 
where $f$ is some function of $S_n$, which is taken to be a real RV for simplicity. Assuming that $S_n$ satisfies an LDP with rate function $I(s)$, we can write
\be
W_n[f]\approx\int_\reals \E^{n[f(s)-I(s)]}\, \D s.
\ee
with sub-exponential corrections in $n$. This integral has the form of a so-called \emp{Laplace integral}, which is known to be dominated for large $n$ by its largest integrand when it is unique. Assuming this is the case, we can proceed to approximate the whole integral as
\be
W_n[f]\approx \E^{n \sup_s [f(s)-I(s)]}.
\ee
Such an approximation is referred to as a \emp{Laplace approximation}, the \emp{Laplace principle} or a \emp{saddle-point approximation} (see Chap.~6 of \cite{bender1978}), and is justified in the context of large deviation theory because the corrections to this approximation are sub-exponential in $n$, as are those of the LDP. By defining the following functional:
\be
\lambda[f]=\lim_{n\ra\infty}\frac{1}{n}\ln W_n(f)
\ee
using a limit similar to the limit defining the LDP, we then obtain
\be
\lambda[f]=\sup_{s\in\reals}\{f(s)-I(s)\}.
\label{eqvaradhan1}
\ee

The result above is what is referred to as \emp{Varadhan Theorem} \cite{varadhan1966,touchette2009}. The contribution of Varadhan was to prove this result for a large class of RVs, which includes not just IID sample means but also random vectors and even random functions, and to rigorously handle all the heuristic approximations used above. As such, Varadhan's Theorem can be considered as a rigorous and general expression of the Laplace principle.

To connect Varadhan's Theorem with the GE Theorem, consider the special case $f(s)=ks$ with $k\in\reals$.\footnote{Varadhan's Theorem holds in its original form for bounded functions $f$, but another stronger version can be applied to unbounded functions and in particular to $f(x)=kx$; see Theorem 1.3.4 of \cite{dupuis1997}.} Then Eq.~(\ref{eqvaradhan1}) becomes
\be
\lambda(k)=\sup_{s\in\reals}\{ks-I(s)\},
\ee
where $\lambda(k)$ is the same function as the one defined in Eq.~(\ref{eqscgf1}). Thus we see that if $S_n$ satisfies an LDP with rate function $I(s)$, then the SCGF $\lambda(k)$ of $S_n$ is the Legendre-Fenchel transform of $I(s)$. This in a sense is the inverse of the GE Theorem, so an obvious question is, can we invert the equation above to obtain $I(s)$ in terms of $\lambda(k)$? The answer provided by the GE Theorem is that, a sufficient condition for $I(s)$ to be the Legendre-Fenchel transform of $\lambda(k)$ is for the latter function to be differentiable.\footnote{The differentiability condition, though sufficient, is not necessary: $I(s)$ in some cases can be the Legendre-Fenchel transform of $\lambda(k)$ even when the latter is not differentiable; see Sec.~4.4 of \cite{touchette2009}.}

This is the closest that we can get to the GE Theorem without proving it. It is important to note, to fully appreciate the importance of that theorem, that it is actually more just than an inversion of Varadhan's Theorem because the existence of $\lambda(k)$ implies the existence of an LDP for $S_n$. In our heuristic inversion of Varadhan's Theorem we assumed that an LDP exists.

\subsection{The contraction principle}

We mentioned before that LDPs can be derived by a contraction method. The basis of this method is the following: let $A_n$ be a random variable $A_n$ known to have an LDP with rate function $I_A(a)$, and consider another random variable $B_n$ which is a function of the first $B_n=f(A_n)$. We want to know whether $B_n$ satisfies an LDP and, if so, to find its rate function.

To find the answer, write the pdf of $B_n$ in terms of the pdf of $A_n$:
\be
p_{B_n}(b)=\int_{\{a: f(a)=b\}} p_{A_n}(a)\, \D a
\ee
and use the LDP for $A_n$ to write
\be
p_{B_n}(b)\approx \int_{\{a: f(a)=b\}} \E^{-nI_A(a)}\, \D a.
\ee
Then apply the Laplace principle to approximate the integral above by its largest term, which corresponds to the minimum of $I(a)$ for  $a$ such that $b=f(a)$. Therefore,
\be
p_{B_n}(b)\approx \exp\left(-n{\inf_{\{a: f(a)=b\}} I_A(a)}\right).
\ee
This shows that $p(B_n)$ also satisfies an LDP with rate function $I_B(b)$ given by
\be
I_B(b)=\inf_{\{a:f(a)=b\}} I_A(a).
\ee

This formula is called the \emp{contraction principle} because $f$ can be many-to-one, i.e., there might be many $a$'s such that $b=f(a)$, in which case we are ``contracting'' information about the rate function of $A_n$ down to $B_n$. In physical terms, this formula is interpreted by saying that an improbable fluctuation\footnote{In statistical physics, a deviation of a random variable away from its typical value is referred to as a \emp{fluctuation}.} of $B_n$ is brought about by the most probable of all improbable fluctuations of $A_n$.

The contraction principle has many applications in statistical physics. In particular, the maximum entropy and minimum free energy principles, which are used to find equilibrium states in the microcanonical and canonical ensembles, respectively, can be seen as deriving from the contraction principle (see Sec.~5 of \cite{touchette2009}).

\subsection{From small to large deviations}

An LDP for a random variable, say $S_n$ again, gives us a lot of information about its pdf. First, we know that $p(S_n)$ concentrates on certain points corresponding to the zeros of the rate function $I(s)$. These points correspond to the most probable or \emp{typical values} of $S_n$ in the limit $n\ra\infty$ and can be shown to be related mathematically to the \emp{Law of Large Numbers} (LLN for short). In fact, an LDP always implies some form of LLN (see Sec.~3.5.7 of \cite{touchette2009}).

Often it is not enough to know that $S_n$ converges in probability to some values; we may also want to determine the likelihood that $S_n$ takes a value away but close to its typical value(s). Consider one such typical value $s^*$ and assume that $I(s)$ admits a Taylor series around $s^*$:
\be
I(s)=I(s^*)+I'(s^*) (s-s^*)+\frac{I''(s^*)}{2} (s-s^*)^2+\cdots.
\ee
Since $s^*$ must correspond to a zero of $I(s)$, the first two terms in this series vanish, and so we are left with the prediction that the \emp{small deviations} of $S_n$ around its typical value are Gaussian-distributed:
\be
p_{S_n}(s)\approx \E^{- n I''(s^*) (s-s^*)^2/2}.
\ee

In this sense, large deviation theory contains the \emp{Central Limit Theorem} (see Sec.~3.5.8 of \cite{touchette2009}). At the same time, large deviation theory can be seen as an extension of the Central Limit Theorem because it gives information not only about the small deviations of $S_n$, but also about its \emp{large deviations} far away from its typical value(s); hence the name of the theory.\footnote{A large deviation is also called a large fluctuation or a \emp{rare and extreme event}.}

\subsection{Exercises}
\label{secex1}

\begin{exercise}
\item \label{excf1}\exdiff{10} (Generating functions) Let $A_n$ be a sum of $n$ IID RVs:
\be
A_n=\sum_{i=1}^n X_i.
\ee 
Show that the generating function of $A_n$, defined as
\be
W_{A_n}(k)=E[\E^{k A_n}],
\ee
satisfies the following factorization property:
\be
W_{A_n}(k)=\prod_{i=1}^n W_{X_i}(k)=W_{X_i}(k)^n,
\ee
$W_{X_i}(k)$ being the generating function of $X_i$. 

\item \label{exgauss1}\exdiff{12} (Gaussian sample mean) Find the expression of $W_X(k)$ for $X$ distributed according to a Gaussian pdf with mean $\mu$ and variance $\sigma^2$. From this result, find $W_{S_n}(k)$ following the previous exercise and $p(S_n)$ by inverse Laplace transform. 

\item \label{exexp1}\exdiff{15} (Exponential sample mean) Repeat the previous exercise for the exponential pdf shown in Eq.~(\ref{eqexpd1}). Obtain from your result the approximation shown in (\ref{eqldexp1}).

\item \label{exbs1}\exdiff{10} (Bernoulli sample mean) Show that the probability distribution of the Bernoulli sample mean is the binomial distribution:
\be
P_{S_n}(s)=\frac{n!}{(sn)![(1-s)n]!}\ \alpha^{sn}(1-\alpha)^{(1-s)n}.
\ee
Use Stirling's approximation to put this result in the form of (\ref{eqldbern1}) with $I(s)$ given by Eq.~(\ref{eqbern1}).

\item \label{exmulti1}\exdiff{12} (Multinomial distribution) Repeat the previous exercise for IID RVs taking values in the set $\{0,1,\ldots, q-1\}$ instead of $\{0,1\}$. Use Stirling's approximation to arrive at an exponential approximation similar to (\ref{eqldbern1}). (See solution in \cite{ellis1999}.)

\item \label{exscgf1}\exdiff{10} (SCGF at the origin) Let $\lambda(k)$ be the SCGF of an IID sample mean $S_n$. Prove the following properties of $\lambda(k)$ at $k=0$:
\begin{itemize}
\item~$\lambda(0)=0$
\item~$\lambda'(0)=E[S_n]=E[X_i]$
\item~$\lambda''(0)=\text{var}(X_i)$
\end{itemize}
Which of these properties remain valid if $S_n$ is not an IID sample mean, but is some general RV?

\item \label{excon1}\exdiff{12} (Convexity of SCGFs) Show that $\lambda(k)$ is a convex function of $k$.

\item \label{exlf1}\exdiff{12} (Legendre transform) Show that, when $\lambda(k)$ is everywhere differentiable and is strictly convex (i.e., has no affine or linear parts), the Legendre-Fenchel transform shown in Eq.~(\ref{eqlf1}) reduces to the \emp{Legendre transform}:
\be
I(s)=k(s) s-\lambda(k(s)),
\ee
where $k(s)$ is the unique of $\lambda'(k)=s$. Explain why the latter equation has a unique root.

\item \label{excon2}\exdiff{20} (Convexity of rate functions) Prove that rate functions obtained from the GE Theorem are strictly convex. 

\item \label{exvar1}\exdiff{12} (Varadhan's Theorem) Verify Varadhan's Theorem for the Gaussian and exponential sample means by explicitly calculating the Legendre-Fenchel transform of $I(s)$ obtained for these sample means. Compare your results with the expressions of $\lambda(k)$ obtained from its definition.

\end{exercise}

\section{Applications of large deviation theory}
\label{secapp}

We study in this section examples of random variables and stochastic processes with interesting large deviation properties. We start by revisiting the simplest application of large deviation theory, namely, IID sample means, and then move to Markov processes, which are often used for modeling physical and man-made systems. More information about applications of large deviation theory in statistical physics can be found in Secs~5 and 6 of \cite{touchette2009}, as well as in the contribution of Engel contained in this volume.

\subsection{IID sample means}
\label{secsanov}

Consider again the sample mean
\be
S_n=\frac{1}{n}\sum_{i=1}^nX_i
\ee
involving $n$ real IID variables $X_1,\ldots,X_n$ with common pdf $p(x)$. To determine whether $S_n$ satisfies an LDP and obtain its rate function, we can follow the GE Theorem and calculate the SCGF $\lambda(k)$ defined in Eq.~(\ref{eqscgf1}). Because of the IID nature of $S_n$, $\lambda(k)$ takes a simple form:
\be
\lambda(k)=\ln E[\E^{kX}].
\ee
In this expression, $X$ is any of the $X_i$'s (remember they are IID). Thus to obtain an LDP for $S_n$, we only have to calculate the log-generating function of $X$, check that it is differentiable and, if so, calculate its Legendre-Fenchel transform (or in this case, its Legendre transform; see Exercise~\ref{exlf1}).  Exercise~\ref{exiid1} considers different distributions $p(x)$ for which this calculation can be carried out, including the Gaussian, exponential and Bernoulli distributions studied before.

As an extra exercise, let us attempt to find the rate function of a special sample mean defined by
\be
L_{n,j}=\frac{1}{n}\sum_{i=1}^n \delta_{X_i,j}
\label{eqempvec1}
\ee
for a sequence $X_1,\ldots,X_n$ of $n$ discrete IID RVs with finite state space $\cX=\{1,2,\ldots,q\}$. This sample mean is called the \emp{empirical distribution} of the $X_i$'s as it ``counts'' the number of times the value or symbol $j\in\cX$ appears in a given realization of $X_1,\ldots,X_n$. This number is normalized by the total number $n$ of RVs, so what we have is the \emp{empirical frequency} for the appearance of the symbol $j$ in realizations of $X_1,\ldots,X_n$.

The values of $L_{n,j}$ for all $j\in\cX$ can be put into a vector $\bL_n$ called the \emp{empirical vector}. For the Bernoulli sample mean, for example, $\cX=\{0,1\}$ and so $\bL_n$ is a two-dimensional vector $\bL_n=(L_{n,0},L_{n,1})$ containing the empirical frequency of $0$'s and $1$'s appearing in a realization $x_1,\ldots,x_n$ of $X_1,\ldots,X_n$:
\be
L_{n,0}=\frac{\#\text{ 0's in } x_1,\ldots,x_n}{n},\quad L_{n,1}=\frac{\#\text{ 1's in } x_1,\ldots,x_n}{n}.
\ee

To find the rate function associated with the random vector $\bL_n$, we apply the GE Theorem but adapt it to the case of random vectors by replacing $k$ in $\lambda(k)$ by a vector $\bk$ having the same dimension as $\bL_n$. The calculation of $\lambda(\bk)$ is left as an exercise (see Exercise~\ref{exsanov1}). The result is
\be
\lambda(\bk)=\ln \sum_{j\in\cX} P_j\, \E^{k_j}
\label{eqsanov1}
\ee
where $P_j=P(X_i=j)$. It is easily checked that $\lambda(\bk)$ is differentiable in the vector sense, so we can use the GE Theorem to conclude that $\bL_n$ satisfies an LDP, which we write as
\be
p(\bL_n={\bmu})\approx \E^{-n I({\bmu})},
\ee
with a vector rate function $I({\bmu})$ given by the Legendre-Fenchel transform of $\lambda(\bk)$. The calculation of this transform is the subject of Exercise~\ref{exsanov1}. The final result is
\be
I({\bmu})=\sum_{j\in\cX}\mu_j\ln\frac{\mu_j}{P_j}.
\label{eqsanov2}
\ee

This rate function is called the \emp{relative entropy} or \emp{Kullback-Leibler divergence} \cite{cover1991}. The full LDP for $\bL_n$ is referred to as \emp{Sanov's Theorem} (see \cite{sanov1961} and Sec.~4.2 of \cite{touchette2009}). 

It can be checked that $I({\bmu})$ is a convex function of $\bmu$ and that it has a unique minimum and zero at ${\bmu}=\mathbf{P}$, i.e., $\mu_j=P_j$ for all $j\in\cX$. As seen before, this property is an expression of the LLN, which says here that the empirical frequencies $L_{n,j}$ converge to the  probabilities $P_j$ as $n\ra\infty$. The LDP goes beyond this result by describing the fluctuations of $\bL_n$ around the typical value ${\bmu}=\mathbf{P}$.

\subsection{Markov chains}
\label{subsecmc1}

Instead of assuming that the sample mean $S_n$ arises from a sum of IID RVs $X_1,\ldots,X_n$, consider the case where the $X_i$'s form a \emp{Markov chain}. This means that the joint pdf $p(X_1,\ldots,X_n)$ has the form
\be
p(X_1,\ldots,X_n)=p(X_1)\prod_{i=1}^{n-1} \pi(X_{i+1}|X_i),
\label{eqmark1}
\ee
where $p(X_i)$ is some initial pdf for $X_1$ and $\pi(X_{i+1}|X_i)$ is the \emp{transition probability density} that $X_{i+1}$ follows $X_i$ in the Markov sequence $X_1\ra X_2\ra\cdots\ra X_n$ (see \cite{grimmett2001} for background information on Markov chains). 

The GE Theorem can still be applied in this case, but the expression of $\lambda(k)$ is more complicated. Skipping the details of the calculation (see Sec.~4.3 of \cite{touchette2009}), we arrive at the following result: provided that the Markov chain is homogeneous and ergodic (see \cite{grimmett2001} for a definition of ergodicity), the SCGF of $S_n$ is given by
\be
\lambda(k)=\ln \zeta(\tPi_k),
\label{eqscgfmc1}
\ee
where $\zeta(\tPi_k)$ is the \emp{dominant eigenvalue} (i.e., with largest magnitude) of the matrix $\tPi_k$ whose elements $\tpi_k(x,x')$ are given by $\tpi_k(x,x')=\pi(x'|x)\E^{k x'}$. We call the matrix $\tPi_k$ the \emp{tilted matrix} associated with $S_n$. If the Markov chain is finite,  it can be proved furthermore that $\lambda(k)$ is analytic and so differentiable. From the GE Theorem, we then conclude that $S_n$ has an LDP with rate function
\be
I(s)=\sup_{k\in\reals}\{ks-\ln \zeta(\tPi_k)\}.
\ee
Some care must be taken if the Markov chain has an infinite number of states. In this case, $\lambda(k)$ is not necessarily analytic and may not even exist (see, e.g., \cite{harris2006,harris2007,rakos2008} for illustrations).

An application of the above result for a simple Bernoulli Markov chain is considered in Exercise~\ref{exbmc1}. In Exercises~\ref{excurr1} and \ref{excurr2}, the case of random variables having the form
\be
Q_n=\frac{1}{n}\sum_{i=1}^{n-1} q(x_i,x_{i+1})
\label{eqcurr1}
\ee
is covered. This type of RV arises in statistical physics when studying currents in stochastic models of interacting particles (see \cite{harris2007}). The tilted matrix $\tPi_k$ associated with $Q_n$ has the form $\tpi_{k}(x,x')=\pi(x'|x)\E^{k q(x,x')}$.

\subsection{Continuous-time Markov processes}

The preceding subsection can be generalized to ergodic Markov processes evolving continuously in time by taking a continuous time limit of discrete-time ergodic Markov chains. 

To illustrate this limit, consider an ergodic continuous-time process described by the state $X(t)$ for $0\leq t \leq T$. For an infinitesimal \emp{time-step} $\Delta t$, time-discretize this process into a sequence of $n+1$ states $X_0,X_1,\ldots,X_n$ with $n=T/\Delta t$ and $X_i=X(i\Delta t)$, $i=0,1,\ldots,n$. The sequence $X_0,X_1,\ldots,X_n$ is an ergodic Markov chain with infinitesimal transition matrix $\Pi(\Delta t)$ given by
\be
\Pi(\Delta t)=\pi (x(t+\Delta t)|x(t))=\E^{G\Delta t}=I+G \Delta t +o(\Delta t),
\ee
where $I$ is the identity matrix and $G$ the \emp{generator} of $X(t)$. With this discretization, it is now possible to study LDPs for processes involving $X(t)$ by studying these processes in discrete time at the level of the Markov chain $X_0,X_1,\ldots,X_n$, and transfer these LDPs into continuous time by taking the limits $\Delta t\ra 0$, $n\ra\infty$. 

As a general application of this procedure, consider a so-called \emp{additive process} defined by 
\be
S_T=\frac{1}{T}\int_0^T X(t)\, \D t.
\ee
The discrete-time version of this RV is the sample mean
\be
S_n=\frac{1}{n\Delta t}\sum_{i=0}^n X_i\, \Delta t=\frac{1}{n}\sum_{i=0}^n X_i.
\ee
From this association, we find that the SCGF of $S_T$, defined by the limit
\be
\lambda(k)=\lim_{T\ra\infty}\frac{1}{T}\ln E[\E^{Tk S_T}],\qquad k\in\reals,
\label{eqscgfct1}
\ee
is given by
\be
\lambda(k)=\lim_{\Delta t\ra 0}\lim_{n\ra\infty}\frac{1}{n \Delta t}\ln E[\E^{k \Delta t\sum_{i=0}^n X_i}]
\ee
at the level of the discrete-time Markov chain. 

According to the previous subsection, the limit above reduces to $\ln\zeta(\tPi_{k}(\Delta t))$, where $\tPi_{k}(\Delta t)$ is the matrix (or operator) corresponding to
\begin{eqnarray}
\tPi_{k}(\Delta t) &=& \E^{k\Delta t x'} \Pi(\Delta t)\nonumber\\
&=& \big(1+kx'\Delta t+o(\Delta t)\big)\big(I+G'\Delta t+o(\Delta t)\big)\nonumber\\
&=& I+\tG_k \Delta t+o(\Delta t),
\label{eqttm1}
\end{eqnarray}
where
\be
\tG_k=G+k x' \delta_{x,x'}.
\ee 
For the continuous-time process $X(t)$, we must therefore have
\be
\lambda(k)=\zeta(\tG_k),
\label{eqscgfmc2}
\ee
where $\zeta(\tG_k)$ is the dominant eigenvalue of the \emp{tilted generator} $\tG_k$.\footnote{As for Markov chains, this result holds for continuous, ergodic processes with finite state-space. For infinite-state or continuous-space processes, a similar result holds provided $G_k$ has an isolated dominant eigenvalue.} Note that the reason why the logarithmic does not appear in the expression of $\lambda(k)$ for the continuous-time process\footnote{Compare with Eq.~(\ref{eqscgfmc1}).} is because $\zeta$ is now the dominant eigenvalue of the generator of the infinitesimal transition matrix, which is itself the exponential of the generator.

With the knowledge of $\lambda(k)$, we can obtain an LDP for $S_T$ using the GE Theorem: If the dominant eigenvalue is differentiable in $k$, then $S_T$ satisfies an LDP in the long-time limit, $T\ra\infty$, which we write as $p_{S_T}(s)\approx \E^{-T I(s)}$, where $I(s)$ is the Legendre-Fenchel transform $\lambda(k)$. Some applications of this result are presented in Exercises~\ref{exrt1} and \ref{excurr2}. Note that in the case of current-type processes having the form of Eq.~(\ref{eqcurr1}), the tilted generator is not simply given by $\tG_k=G+kx'\delta_{x,x'}$; see Exercise~\ref{excurr2}.

\subsection{Paths large deviations}
\label{subsecpathldp}

We complete our tour of mathematical applications of large deviation theory by studying a different large deviation limit, namely, the low-noise limit of the following \emp{stochastic differential equation} (SDE for short):
\be
\dot x(t)=f(x(t))+\sqrt{\vep}\, \xi(t),\qquad x(0)=0,
\label{eqsde1}
\ee 
which involves a \emp{force} $f(x)$ and a \emp{Gaussian white noise} $\xi(t)$ with the properties $E[\xi(t)]=0$ and $E[\xi(t')\xi(t)]=\delta(t'-t)$; see Chap.~XVI of \cite{kampen1992} for background information on SDEs.

We are interested in studying for this process the pdf of a given \emp{random path} $\{x(t)\}_{t=0}^T$ of duration $T$ in the limit where the \emp{noise power} $\vep$ vanishes. This abstract pdf can be defined heuristically using path integral methods (see Sec.~6.1 of \cite{touchette2009}). In the following, we denote this pdf by the functional notation $p[x]$ as a shorthand for $p(\{x(t)\}_{t=0}^T)$.

The idea behind seeing the low-noise limit as a large deviation limit is that, as $\vep\ra 0$, the random path arising from the SDE above should converge in probability to the \emp{deterministic path} $x(t)$ solving the ordinary differential equation
\be
\dot x(t)=f(x(t)),\qquad x(0)=0.
\label{eqdet1}
\ee
This convergence is a LLN-type result, and so in the spirit of large deviation theory we are interested in quantifying the likelihood that a random path $\{x(t)\}_{t=0}^T$ ventures away from the deterministic path in the limit $\vep\ra 0$. The functional LDP that characterizes these path fluctuations has the form
\be
p[x]\approx \E^{-I[x]/\vep},
\ee
where
\be
I[x]=\int_0^T [\dot x(t)-f(x(t))]^2\, \D t.
\label{eqldpf1}
\ee
See \cite{freidlin1984} and Sec.~6.1 of \cite{touchette2009} for historical sources on this LDP.

The rate functional $I[x]$ is called the \emp{action}, \emp{Lagrangian} or \emp{entropy} of the path $\{x(t)\}_{t=0}^T$. The names ``action'' and ``Lagrangian'' come from an analogy with the action of quantum trajectories in the path integral approach of quantum mechanics (see Sec.~6.1 of \cite{touchette2009}). There is also a close analogy between the low-noise limit of SDEs and the semi-classical or WKB approximation of quantum mechanics \cite{touchette2009}. 

The path LDP above can be generalized to higher-dimensional SDEs as well as SDEs involving state-dependent noise and correlated noises (see Sec.~6.1 of \cite{touchette2009}). In all cases, the minimum and zero of the rate functional is the trajectory of the deterministic system obtained in the zero-noise limit. This is verified for the 1D system considered above: $I[x]\geq 0$ for all trajectories and $I[x]=0$ for the unique trajectory solving Eq.~(\ref{eqdet1}).

Functional LDPs are the most refined LDPs that can be derived for SDEs as they characterize the probability of complete trajectories. Other ``coarser'' LDPs can be derived from these by contraction. For example, we might be interested to determine the pdf $p(x,T)$ of the state $x(T)$ reached after a time $T$. The contraction in this case is obvious: $p(x,T)$ must have the large deviation form $p(x,T)\approx \E^{-V(x,T)/\vep}$ with
\be
V(x,T)=\inf_{x(t):x(0)=0,x(T)=x} I[x].
\ee
That is, the probability of reaching $x(T)=x$ from $x(0)=0$ is determined by the path connecting these two endpoints having the largest probability. We call this path the \emp{optimal path}, \emp{maximum likelihood path} or \emp{instanton}. Using variational calculus techniques, often used in classical mechanics, it can be proved that this path satisfies an Euler-Lagrange-type equation as well as a Hamilton-type equation (see Sec.~6.1 of \cite{touchette2009}). Applications of these equations are covered in Exercises~\ref{exoup1}--\ref{expol1}.

Quantities similar to the additive process $S_T$ considered in the previous subsection can also be defined for SDEs (see Exercises~\ref{exrt1} and \ref{excurr1}). An interesting aspect of these quantities is that their LDPs can involve the noise power $\vep$ if the low-noise limit is taken, as well as the integration time $T$, which arises because of the additive (in the sense of sample mean) nature of these quantities. In this case, the limit $T\ra 0$ must be taken before the low-noise limit $\vep\ra 0$. If $\vep\ra 0$ is taken first, the system considered is no longer random, which means that there can be no LDP in time.

\subsection{Physical applications}

Some applications of the large deviation results seen so far are covered in the contribution of Engel to this volume. The following list gives an indication of these applications and some key references for learning more about them. A more complete presentation of the applications of large deviation theory in statistical physics can be found in Secs.~5 and 6 of ~\cite{touchette2009}.
\begin{itemize}
\item \emp{Equilibrium systems:} Equilibrium statistical mechanics, as embodied by the ensemble theory of Boltzmann and Gibbs, can be seen with hindsight as a large deviation theory of many-body systems at equilibrium. This becomes evident by realizing that the thermodynamic limit is a large deviation limit, that the entropy is the equivalent of a rate function and the free energy the equivalent of a SCGF. Moreover, the Legendre transform of thermodynamics connecting the entropy and free energy is nothing but the Legendre-Fenchel transform connecting the rate function and the SCGF in the GE Theorem and in Varadhan's Theorem. For a more complete explanation of these analogies, and historical sources on the development of large deviation theory in relation to equilibrium statistical mechanics, see the book by Ellis \cite{ellis1985} and Sec.~3.7 of \cite{touchette2009}. 
 
\item \emp{Chaotic systems and multifractals:} The so-called \emp{thermodynamic formalism} of dynamical systems, developed by Ruelle \cite{ruelle2004} and others, can also be re-interpreted with hindsight as an application of large deviation theory for the study of chaotic systems.\footnote{There is a reference to this connection in a note of Ruelle's popular book, Chance and Chaos \cite{ruelle1993} (see~Note 2 of Chap.~19).} There are two quantities in this theory playing the role of the SCGF, namely, the \emp{topological pressure} and the \emp{structure function}. The Legendre transform appearing in this theory is also an analogue of the one encountered in large deviation theory. References to these analogies can be found in Secs.~7.1 and 7.2 of \cite{touchette2009}.

\item \emp{Nonequilibrium systems:} Large deviation theory is becoming the standard formalism used in studies of nonequilibrium systems modelled by SDEs and Markov processes in general. In fact, large deviation theory is currently experiencing a sort of revival in physics and mathematics as a result of its growing application in nonequilibrium statistical mechanics. Many LDPs have been derived in this context: LDPs for the current fluctuations or the occupation of interacting particle models, such as the exclusion process, the zero-range process (see \cite{harris2005}) and their many variants \cite{spohn1991}, as well as LDPs for work-related and entropy-production-related quantities for nonequilibrium systems modelled with SDEs \cite{chetrite2008,chetrite2007a}. Good entry points in this vast field of research are \cite{derrida2007} and \cite{harris2007}. 

\item \emp{Fluctuation relations:} Several LDPs have come to be studied in recent years under the name of fluctuation relations. To illustrate these results, consider the additive process $S_T$ considered earlier and assume that this process admits an LDP with rate function $I(s)$. In many cases, it is interesting not only to know that $S_T$ admits an LDP but to know how probable positive fluctuations of $S_T$ are compared to negative fluctuations. For this purpose, it is common to study the ratio
\be
\frac{p_{S_T}(s)}{p_{S_T}(-s)},
\ee
which reduces to
\be
\frac{p_{S_T}(s)}{p_{S_T}(-s)}\approx \E^{T[I(-s)-I(s)]}
\ee
if we assume an LDP for $S_T$. In many cases, the difference $I(-s)-I(s)$ is linear in $s$, and one then says that $S_T$ satisfies a \emp{conventional fluctuation relation}, whereas if it nonlinear in $s$, then one says that $S_T$ satisfies an \emp{extended fluctuation relation}. Other types of fluctuation relations have come to be defined in addition to these; for more information, the reader is referred to Sec.~6.3 of \cite{touchette2009}. A list of the many physical systems for which fluctuation relations have been derived or observed can also be found in this reference.
\end{itemize}

\subsection{Exercises}
\label{secex2}

\begin{exercise}
\item \label{exiid1}\exdiff{12} (IID sample means) Use the GE Theorem to find the rate function of the IID sample mean $S_n$ for the following probability distribution and densities:
\begin{itemize}
\item~Gaussian:
\be
p(x)=\frac{1}{\sqrt{2\pi \sigma^2}} \E^{-(x-\mu)^2/(2\sigma^2)},\qquad x\in\reals.
\ee
\item~Bernoulli: $X_i\in\{0,1\}$, $P(X_i=0)=1-\alpha$, $P(X_i=1)=\alpha$.
\item~Exponential: 
\be
p(x)=\frac{1}{\mu} \E^{-x/\mu},\qquad x\in [0,\infty),\quad \mu>0.
\ee
\item~Uniform:
\be
p(x)=\left\{
\begin{array}{lll}
\frac{1}{a} & & x\in[0,a]\\
0 & & \text{otherwise}.
\end{array}
\right.
\ee
\item~Cauchy:
\be
p(x)=\frac{\sigma}{\pi(x^2+\sigma^2)},\qquad x\in\reals,\quad \sigma>0.
\ee
\end{itemize}

\item \label{exnc1}\exdiff{15} (Nonconvex rate functions) Rate functions obtained from the GE Theorem are necessarily convex (strictly convex, in fact; see Exercise~\ref{excon2}), but rate functions in general need not be convex. Consider, as an example, the process defined by
\be
S_n=Y+\frac{1}{n}\sum_{i=1}^n X_i,
\ee
where $Y=\pm1$ with probability $\frac{1}{2}$ and the $X_i$'s are Gaussian IID RVs. Find the rate function for $S_n$ assuming that $Y$ is independent of the $X_i$'s. Then find the corresponding SCGF. What is the relation between the Legendre-Fenchel transform of $\lambda(k)$ and the rate function $I(s)$? How is the nonconvexity of $I(s)$ related to the differentiability of $\lambda(k)$? (See Example~4.7 of \cite{touchette2009} for the solution.)

\item \label{exexp2}\exdiff{15} (Exponential mixture) Repeat the previous exercise by replacing the Bernoulli $Y$ with $Y=Z/n$, where $Z$ is an exponential RV with mean $1$. (See Example~4.8 of \cite{touchette2009} for the solution.)

\item \label{exprod1}\exdiff{12} (Product process) Find the rate function of 
\be
Z_n=\left(\prod_{i=1}^n X_i\right)^{1/n}
\ee
where $X_i\in\{1,2\}$ with $p(X_i=1)=\alpha$ and $p(X_i=2)=1-\alpha$, $0<\alpha<1$.

\item \label{exsp1}\exdiff{15} (Self-process) Consider a sequence of IID Gaussian RVs $X_1,\ldots,X_n$. Find the rate function of the so-called \emp{self-process} defined by 
\be
S_n=-\frac{1}{n}\ln p(X_1,\ldots,X_n).
\ee
(See Sec.~4.5 of \cite{touchette2009} for information about this process.) Repeat the problem for other choices of pdf for the RVs.

\item \label{exild1}\exdiff{20} (Iterated large deviations~\cite{lowe1996}) Consider the sample mean
\be
S_{m,n}=\frac{1}{m}\sum_{i=1}^m X_{i}^{(n)}
\ee
involving $m$ IID copies of a random variable $X^{(n)}$. Show that if $X^{(n)}$ satisfies an LDP of the form
\be
p_{X^{(n)}}(x)\approx \E^{-nI(x)},
\ee
then $S_{m,n}$ satisfies an LDP having the form
\be
p_{S_{m,n}}(s)\approx \E^{-mn I(x)}.
\ee

\item \label{exprw1}\exdiff{40} (Persistent random walk) Use the result of the previous exercise to find the rate function of 
\be
S_n=\frac{1}{n}\sum_{i=1}^n X_i,
\ee
where the $X_i$'s are independent Bernoulli RVs with non-identical distribution $P(X_i=0)=\alpha^i$ and $P(X_i=1)=1-\alpha^i$ with $0<\alpha<1$.

\item \label{exsanov1}\exdiff{15} (Sanov's Theorem) Consider the empirical vector $\bL_n$ with components defined in Eq.~(\ref{eqempvec1}). Derive the  expression found in Eq.~(\ref{eqsanov1}) for
\be
\lambda(\bk)=\lim_{n\ra\infty}\frac{1}{n}\ln E[\E^{n\, \bk\cdot\bL_n}],
\ee 
where $\bk\cdot\bL_n$ denotes the scalar product of $\bk$ and $\bL_n$. Then obtain the rate function found in (\ref{eqsanov2}) by calculating the Legendre transform of $\lambda(\bk)$. Check explicitly that the rate function is convex and has a single minimum and zero.

\item \label{excp1}\exdiff{20} (Contraction principle) Repeat the first exercise of this section by using the contraction principle. That is, use the rate function of the empirical vector $\bL_n$ to obtain the rate function of $S_n$. What is the mapping from $\bL_n$ to $S_n$? Is this mapping many-to-one or one-to-one?
 
\item \label{exbmc1}\exdiff{17} (Bernoulli Markov chain) Consider the Bernoulli sample mean 
\be
S_n=\frac{1}{n}\sum_{i=1}^n X_i,\qquad X_i\in\{0,1\}.
\ee
Find the expression of $\lambda(k)$ and $I(s)$ for this process assuming that the $X_i$'s form a Markov process with symmetric transition matrix 
\be
\pi (x'|x)=\left\{
\begin{array}{lll}
1-\alpha & & x'=x\\
\alpha & & x'\neq x
\end{array}
\right.
\ee
with $0\leq \alpha\leq 1$. (See Example~4.4 of \cite{touchette2009} for the solution.)

\item \label{exrt1}\exdiff{20} (Residence time) Consider a Markov process with state space $\cX=\{0,1\}$ and generator
\be
G=\left(
\begin{array}{cc}
-\alpha & \beta\\
\alpha & -\beta
\end{array}
\right).
\ee
Find for this process the rate function of the random variable
\be
L_T=\frac{1}{T} \int_0^T \delta_{X(t),0}\, \D t,
\ee 
which represents the fraction of time the state $X(t)$ of the Markov chain spends in the state $X=0$ over a period of time $T$.

\item \label{excurr1}\exdiff{20} (Current fluctuations in discrete time) Consider the Markov chain of Exercise~\ref{exbmc1}. Find the rate function of the \emp{mean current} $Q_n$ defined as 
\be
Q_n=\frac{1}{n}\sum_{i=1}^n f(x_{i+1},x_i),
\ee 
\be
f (x',x)=\left\{
\begin{array}{lll}
0 & & x'=x\\
1 & & x'\neq x.
\end{array}
\right.
\ee
$Q_n$ represents the mean number of jumps between the states $0$ and $1$: at each transition of the Markov chain, the current is incremented by 1 whenever a jump between the states $0$ and $1$ occurs.

\item \label{excurr2}\exdiff{22} (Current fluctuations in continuous time) Repeat the previous exercise for the Markov process of Exercise~\ref{exrt1} and the current $Q_T$ defined as
\be
Q_T=\lim_{\Delta t\ra 0}\frac{1}{T}\sum_{i=0}^{n-1} f(x_{i+1},x_i),
\ee
where $x_i=x(i\Delta t)$ ad $f(x',x)$ as above, i.e., $f(x_{i+1},x_i)=1-\delta_{x_i,x_{i+1}}$. Show for this process that the tilted generator is
\be
\tG_k=\left(
\begin{array}{cc}
-\alpha & \beta\,\E^k\\
\alpha\, \E^k & -\beta
\end{array}
\right).
\ee

\item \label{exoup1}\exdiff{17} (Ornstein-Uhlenbeck process) Consider the following linear SDE:
\be
\dot x(t)=-\gamma x(t)+\sqrt{\vep}\, \xi(t),
\ee
often referred to as the \emp{Langevin equation} or \emp{Ornstein-Uhlenbeck process}. Find for this SDE the solution of the optimal path connecting the initial point $x(0)=0$ with the fluctuation $x(T)=x$. From the solution, find the pdf $p(x,T)$ as well as the stationary pdf obtained in the limit $T\ra\infty$. Assume $\vep\ra 0$ throughout but then verify that the results obtained are valid for all $\vep>0$.

\item \label{excons1}\exdiff{20} (Conservative system) Show by large deviation techniques that the stationary pdf of the SDE 
\be
\dot \bx (t)=-\nabla U(\bx(t))+\sqrt{\vep}\, \xi(t)
\ee
admits the LDP $p(\bx)\approx \E^{-V(\bx)/\vep}$ with rate function $V(\bx)=2U(\bx)$. The rate function $V(\bx)$ is called the \emp{quasi-potential}.

\item \label{extrans1}\exdiff{25} (Transversal force) Show that the LDP found in the previous exercise also holds for the SDE
\be
\dot \bx (t)=-\nabla U(\bx(t))+\mathbf{A}(\bx)+\sqrt{\vep}\, \xi(t)
\ee
if $\nabla U\cdot \mathbf{A}=0$. (See Sec.~4.3 of \cite{freidlin1984} for the solution.)

\item \label{expol1}\exdiff{25} (Noisy Van der Pol oscillator) The following set of SDEs describes a noisy version of the well-known Van der Pol equation:
\begin{eqnarray}
\dot x&=&v\nonumber\\
\dot v&=&-x+v(\alpha-x^2-v^2)+\sqrt{\vep}\, \xi.
\end{eqnarray}
The parameter $\alpha\in\reals$ controls a bifurcation of the zero-noise system: for $\alpha\leq 0$, the system with $\xi=0$ has a single attracting point at the origin, whereas for $\alpha>0$, it has a stable limit cycle centered on the origin. Show that the stationary distribution of the noisy oscillator has the large deviation form $p(x,v)\approx \E^{-U(x,v)/\vep}$ as $\vep\ra 0$ with rate function 
\be
U(x,v)=-\alpha (x^2+v^2)+\frac{1}{2}(x^2+v^2)^2.
\ee
Find a different set of SDEs that has the same rate function. (See \cite{graham1989} for the solution.)

\item \label{exbmp1}\exdiff{30} (Dragged Brownian particle \cite{zon2003a}) The stochastic dynamics of a tiny glass bead immersed in water and pulled with laser tweezers at constant velocity can be modelled in the overdamped limit by the following reduced Langevin equation:
\be
\dot x=-(x(t)-\nu t)+\sqrt{\vep}\, \xi(t),
\ee
where $x(t)$ is the position of the glass bead, $\nu$ the pulling velocity, $\xi(t)$ a Gaussian white noise modeling the influence of the surrounding fluid, and $\vep$ the noise power related to the temperature of the fluid. Show that the \emp{mean work} done by the laser as it pulls the glass bead over a time $T$, which is defined as
\be
W_T=-\frac{\nu}{T}\int_0^T (x(t)-\nu t)\, \D t,
\ee
satisfies an LDP in the limit $T\ra\infty$. Derive this LDP by assuming first that $\vep\ra 0$, and then derive the LDP without this assumption. Finally, determine whether $W_T$ satisfies a conventional or extended fluctuation relation, and study the limit $\nu\ra 0$. (See Example~6.9 of \cite{touchette2009} for more references on this model.)
\end{exercise}

\section{Numerical estimation of large deviation probabilities}
\label{secnum1}

The previous sections might give the false impression that rate functions can be calculated explicitly for many stochastic processes. In fact, exact and explicit expressions of rate functions can be found only in few simple cases. For most stochastic processes of scientific interest (e.g,, noisy nonlinear dynamical systems, chemical reactions, queues, etc.), we have to rely on analytical approximations and numerical methods to evaluate rate functions.

The rest of these notes attempts to give an overview of some of these numerical methods and to illustrate them with the simple processes (IID sample means and Markov processes) treated before. The goal is to give a flavor of the general ideas behind large deviation simulations rather than a complete survey, so only a subset of the many methods that have come to be devised for numerically estimating rate functions are covered in what follows. Other methods not treated here, such as the transition path method discussed by Dellago in this volume, are mentioned at the end of the next section.

\subsection{Direct sampling}

The problem addressed in this section is to obtain a numerical estimate of the pdf $p_{S_n}(s)$ for a real random variable $S_n$ satisfying an LDP, and to extract from this an estimate of the rate function $I(s)$.\footnote{The literature on large deviation simulation usually considers the estimation of $P(S_n\in A)$ for some set $A$ rather than the estimation of the whole pdf of $S_n$.} To be general, we take $S_n$ to be a function of $n$ RVs $X_1,\ldots,X_n$, which at this point are not necessarily IID. To simplify the notation, we will use the shorthand $\bom=X_1,\ldots,X_n$. Thus we write $S_n$ as the function $S_n(\bom)$ and denote by $p(\bom)$ the joint pdf of the $X_i$'s.\footnote{For simplicity, we consider the $X_i$'s to be real RVs. The case of discrete RVs follow with slight changes of notations.}

Numerically, we cannot of course obtain $p_{S_n}(s)$ or $I(s)$ for all $s\in\reals$, but only for a finite number of values $s$, which we take for simplicity to be equally spaced with a small step $\Delta s$. Following our discussion of the LDP, we thus attempt to estimate the coarse-grained pdf
\be
p_{S_n}(s)=\frac{P(S_n\in [s,s+\Delta s])}{\Delta s}=\frac{P(S_n\in\Delta_s)}{\Delta s},
\label{eqcgpdf1}
\ee 
where $\Delta_s$ denotes the small interval $[s,s+\Delta s]$ anchored at the value $s$.\footnote{Though not explicitly noted, the coarse-grained pdf depends on $\Delta s$; see Footnote~\ref{fnldp1}.} 

To construct this estimate, we follow the \emp{statistical sampling} or \emp{Monte Carlo method}, which we broke down into the following steps (see the contribution of Katzgraber in this volume for more details):
\begin{enumerate}
\item Generate a \emp{sample} $\{\bom^{(j)}\}_{j=1}^L$ of $L$ copies or realizations of the sequence $\bom$ from its pdf $p(\bom)$.
\item Obtain from this sample a sample $\{s^{(j)}\}_{j=1}^L$ of values or realizations for $S_n$:
\be
s^{(j)}=S_n(\bom^{(j)}),\qquad j=1,\ldots,L.
\ee
\item Estimate $P(S_n\in\Delta_s)$ by calculating the sample mean
\be
\hat P_L(\Delta_s)=\frac{1}{L}\sum_{j=1}^L \id_{\Delta_s}(s^{(j)}),
\ee
where $\id_A(x)$ denotes the \emp{indicator function} for the set $A$, which is equal to $1$ if $x\in A$ and $0$ otherwise. 

\item Turn the \emp{estimator} $\hat P_L(\Delta_s)$ of the probability $P(S_n\in\Delta_s)$ into an estimator $\hp_L(s)$ of the probability density $p_{S_n}(s)$:
\be
\hp_L(s)=\frac{\hat P_L(\Delta_s)}{\Delta s}=\frac{1}{L\Delta s}\sum_{j=1}^L \id_{\Delta_s}(s^{(j)}).
\ee
\end{enumerate}

\begin{figure}[t]
\resizebox{\textwidth}{!}{\includegraphics{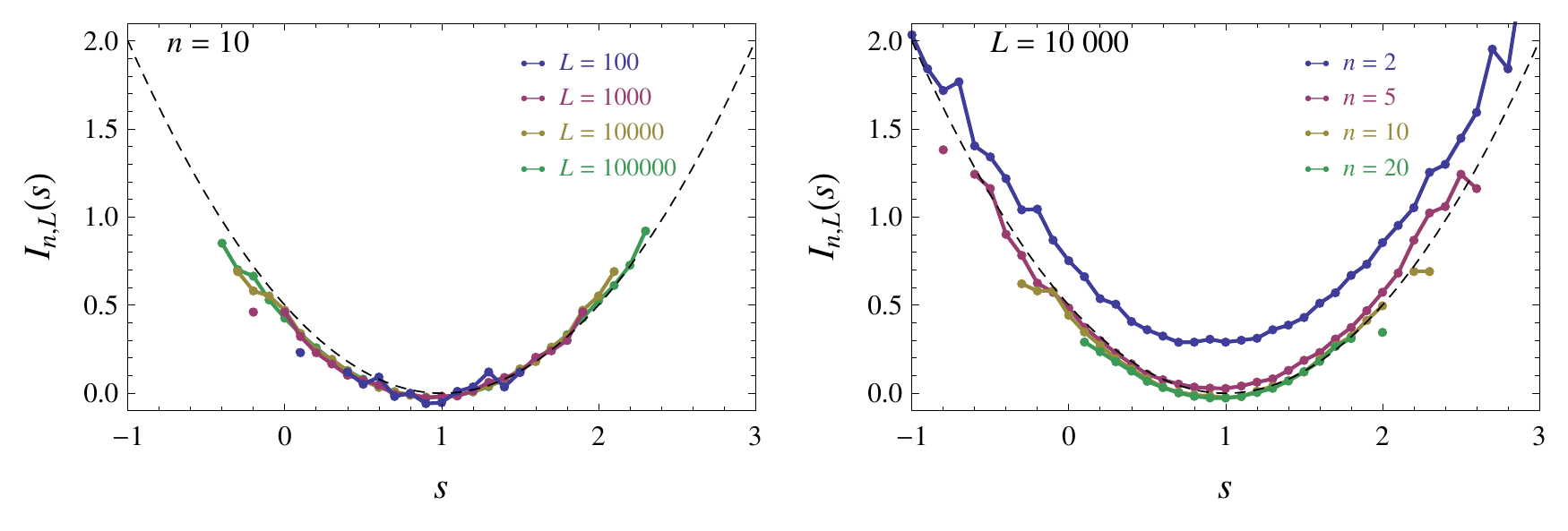}}
\caption{(Left) Naive sampling of the Gaussian IID sample mean ($\mu=\sigma=1$) for $n=10$ and different sample sizes $L$. The dashed line is the exact rate function. As $L$ grows, a larger range of $I(s)$ is sampled. (Right) Naive sampling for a fixed sample size $L=10\, 000$ and various values of $n$. As $n$ increases, $I_{n,L}(s)$ approaches the expected rate function but the sampling becomes inefficient as it becomes restricted to a narrow domain.}
\label{figdirsamp1}
\end{figure}

The result of these steps is illustrated in Fig.~\ref{figdirsamp1} for the case of the IID Gaussian sample mean. Note that $\hp_L(s)$ above is nothing but an empirical vector for $S_n$ (see Sec.~\ref{secsanov}) or a \emp{density histogram} of the sample $\{s^{(j)}\}_{j=1}^L$.

The reason for choosing $\hp(s)$ as our estimator of $p_{S_n}(s)$ is that it is an \emp{unbiased estimator} in the sense that
\be
E[\hp_L(s)]=p_{S_n}(s)
\ee
for all $L$. Moreover, we know from the LLN that $\hp_L(s)$ converges in probability to its mean $p_{S_n}(s)$ as $L\ra\infty$. Therefore, the larger our sample, the closer we should get to a valid estimation of $p_{S_n}(s)$. 

To extract a rate function $I(s)$ from $\hp_L(s)$, we simply compute
\be
I_{n,L}(s)=-\frac{1}{n}\ln \hp_L(s)
\label{eqrfe1}
\ee
and repeat the whole process for larger and larger integer values of $n$ and $L$ until $I_{n,L}(s)$ converges to some desired level of accuracy.\footnote{Note that $\hat P_L(\Delta_s)$ and $\hp_L(s)$ differ only by the factor $\Delta s$. $I_{n,L}(s)$ can therefore be computed from either estimator with a difference $(\ln \Delta s)/n$ that vanishes as $n\ra\infty$.}

\subsection{Importance sampling}

A basic rule of thumb in statistical sampling, suggested by the LLN, is that an event with probability $P$ will appear in a sample of size $L$ roughly $LP$ times. Thus to get at least one instance of that event in the sample, we must have $L>1/P$ as an approximate lower bound for the size of our sample (see Exercise~\ref{exerrest1} for a more precise derivation of this estimate).

Applying this result to $p_{S_n}(s)$, we see that, if this pdf satisfies an LDP of the form $p_{S_n}(s)\approx \E^{-nI(s)}$, then we need to have $L>\E^{n I(s)}$ to get at least one instance of the event $S_n\in \Delta_s$ in our sample. In other words, our sample must be exponentially large with $n$ in order to see any large deviations (see Fig.~\ref{figdirsamp1} and Exercise~\ref{exds1}).

This is a severe limitation of the sampling scheme outlined earlier, and we call it for this reason \emp{crude Monte Carlo} or \emp{naive sampling}. The way around this limitation is to use \emp{importance sampling} (IS for short) which works basically as follows (see the contribution of Katzgraber for more details):
\begin{enumerate}
\item Instead of sampling the $X_i$'s according to the joint pdf $p(\bom)$, sample them according to a new pdf $q (\bom)$;\footnote{The pdf $q$ must have a support at least as large as that of $p$, i.e., $q(\bom)>0$ if $p(\bom)>0$, otherwise the ratio $R_n$ is ill-defined. In this case, we say that $q$ is \emp{relatively continuous} with respect to $p$ and write $q\gg p$.}
\item Calculate instead of $\hp_L(s)$ the estimator
\be
\hq_L(s)=\frac{1}{L\Delta s}\sum_{j=1}^L \id_{\Delta_s} \big(S_n(\bom^{(j)})\big)\, R(\bom^{(j)}),
\label{eqqest1}
\ee
where 
\be
R(\bom)=\frac{p(\bom)}{q(\bom)}
\ee
is the called the \emp{likelihood ratio}. Mathematically, $R$ also corresponds to the \emp{Radon-Nikodym derivative} of the measures associated with $p$ and $q$.\footnote{The Radon-Nikodym derivative of two measures $\mu$ and $\nu$ such that $\nu\gg\mu$ is defined as $d\mu/d\nu$. If these measures have densities $p$ and $q$, respectively, then $d\mu/d\nu=p/q$.}
\end{enumerate}

The new estimator $\hq_L(s)$ is also an unbiased estimator of $p_{S_n}(s)$ because (see Exercise~\ref{exuie1})
\be
E_q[\hq_L(s)]= E_p[\hp_L(s)].
\label{equb1}
\ee
However, there is a reason for choosing $\hq_L(s)$ as our new estimator: we might be able to come up with a suitable choice for the pdf $q$ such that $\hq_L(s)$ has a smaller variance than $\hp_L(s)$. This is in fact the goal of IS: select $q(\bom)$ so as to minimize the variance
\be
\var_q(\hq_L(s))=E_q[(\hq_L(s)-p(s))^2].
\ee
If we can choose a $q$ such that $\var_q(\hq_L(s))<\var_p(\hp_L(s))$, then $\hq_L(s)$ will converge faster to $p_{S_n}(s)$ than $\hp_L(s)$ as we increase $L$.

It can be proved (see Exercise~\ref{exoptq1}) that in the class of all pdfs that are relatively continuous with respect to $p$ there is a unique pdf $q^*$ that minimizes the variance above. This \emp{optimal} IS pdf has the form
\be
q^*(\bom) =\id_{\Delta_s}(S_n(\bom))\frac{p(\bom)}{p_{S_n}(s)} =p(\bom|S_n(\bom)=s)
\label{eqoippdf1}
\ee
and has the desired property that $\var_{q^*}(\hq_L(s))=0$. It does seem therefore that our sampling problem is solved, until one realizes that $q^*$ involves the unknown pdf that we want to estimate, namely, $p_{S_n}(s)$. Consequently, $q^*$ cannot be used in practice as a sampling pdf. Other pdfs must be considered.

The next subsections present a more practical sampling pdf, which can be proved to be optimal in the large deviation limit $n\ra\infty$. We will not attempt to justify the form of this pdf nor will we prove its optimality. Rather, we will attempt to illustrate how it works in the simplest way possible by applying it to IID sample means and Markov processes. For a more complete and precise treatment of IS of large deviation probabilities, see \cite{bucklew2004} and Chap.~VI of \cite{asmussen2007}. For background information on IS, see Katzgraber (this volume) and Chap.~V of \cite{asmussen2007}.

\subsection{Exponential change of measure}

An important class of IS pdfs used for estimating $p_{S_n}(s)$ is given by
\be
p_k(\bom)=\frac{\E^{nk S_n(\bom)}}{W_n(k)}p(\bom),
\ee
where $k\in\reals$ and $W_n(k)$ is a normalization factor given by
\be
W_n(k)=E_p[\E^{nkS_n}]=\int_{\reals^n}\E^{nkS_n(\bom)}\, p(\bom)\, \D\bom.
\ee
We call this class or family of pdfs parameterized by $k$ the \emp{exponential family}. In large deviation theory, $p_k$ is also known as the \emp{tilted pdf} associated with $p$ or the \emp{exponential twisting} of $p$.\footnote{In statistics and actuarial mathematics, $p_k$ is known as the \emp{associated law} or \emp{Esscher transform} of $p$.} The likelihood ratio associated with this change of pdf is 
\be
R(\bom)=\E^{-nkS_n(\bom)}\, W_n(k),
\ee
which, for the purpose of estimating $I(s)$, can be approximated by
\be
R(\bom)\approx \E^{-n[k S_n(\bom)-\lambda(k)]}
\ee
given the large deviation approximation $W_n(k)\approx\E^{n\lambda(k)}$.

It is important to note that a single pdf of the exponential family with parameter $k$ cannot be used to efficiently sample the whole of the function $p_{S_n}(s)$ but only a particular point of that pdf. More precisely, if we want to sample $p_{S_N}(s)$ at the value $s$, we must choose $k$ such that
\be
E_{p_k}[\hq_L(s)]=p_{S_n}(s)
\label{eqecm1}
\ee
which is equivalent to solving $\lambda'(k)=s$ for $k$, where
\label{eqscgfis1}
\be
\lambda(k)=\lim_{n\ra\infty}\frac{1}{n}\ln E_p[\E^{nk S_n}]
\ee
is the SCGF of $S_n$ defined with respect to $p$ (see Exercise~\ref{exisscgf1}). Let us denote the solution of these equations by $k(s)$. For many processes $S_n$, it can be proved that the tilted pdf $p_{k(s)}$ corresponding to $k(s)$ is asymptotically optimal in $n$, in the sense that the variance of $\hq_L(s)$ under $p_{k(s)}$ goes asymptotically to $0$ as $n\ra\infty$. When this happens, the convergence of $\hq_L(s)$ towards $p_{S_n}(s)$ is fast for increasing $L$ and requires a sub-exponential number of samples to achieve a given accuracy for $I_{n,L}(s)$. To obtain the full rate function, we then simply need to repeat the sampling for other values of $s$, and so other values of $k$, and scale the whole process for larger values of $n$.

In practice, it is often easier to reverse the roles of $k$ and $s$ in the sampling. That is, instead of fixing $s$ and selecting $k=k(s)$ for the numerical estimation of $I(s)$, we can fix $k$ to obtain the rate function at $s(k)=\lambda'(k)$. This way, we can build a parametric representation of $I(s)$ over a certain range of $s$ values by covering or ``scanning'' enough values of $k$.

At this point, there should be a suspicion that we will not achieve much with this \emp{exponential change of measure} method (ECM method short) since the forms of $p_k$ and $\hq_L(s)$ presuppose the knowledge of $W_n(k)$ and in turn $\lambda(k)$. We know from the GE Theorem that $\lambda(k)$ is in many cases sufficient to obtain $I(s)$, so why taking the trouble of sampling $\hq_L(s)$ to get $I(s)$?\footnote{To make matters worst, the sampling of $\hq_L(s)$ does involve $I(s)$ directly; see Exercise~\ref{exisi1}.}

The answer to this question is that $I(s)$ can be estimated from $p_{k}$ without knowing $W_n(k)$ using two insights:
\begin{itemize} 
\item Estimate $I(s)$ indirectly by sampling an estimator of $\lambda(k)$ or of $S_n$, which does not involve $W_n(k)$, instead of sampling the estimator $\hq_L(s)$ of $p_{S_n}(s)$;
\item Sample $\bom$ according to the a priori pdf $p(\bom)$ or use the Metropolis algorithm, also known as the Markov chain Monte Carlo algorithm, to sample $\bom$ according to $p_k(\bom)$ without knowing $W_n(k)$ (see Appendix~\ref{appmh} and the contribution of Katzgraber in this volume).
\end{itemize}
These points will be explained in the next section (see also Exercise~\ref{exmetis1}).

For the rest of this subsection, we will illustrate the ECM method for a simple Gaussian IID sample mean, leaving aside the issue of having to know $W_n(k)$. For this process 
\be
p_k(\bom)=p_k(x_1,\ldots,x_n)=\prod_{i=1}^n p_k(x_i),
\ee
where
\be
p_k(x_i)=\frac{\E^{k x_i}\, p(x_i)}{W(k)},\qquad W(k)=E_p[\E^{k X}],
\ee
with $p(x_i)$ a Gaussian pdf with mean $\mu$ and variance $\sigma^2$. We know from Exercise~\ref{exiid1} the expression of the generating function $W(k)$ for this pdf:
\be
W(k)=\E^{k\mu+\frac{\sigma^2}{2}k^2}.
\ee
The explicit expression of $p_k(x_i)$ is therefore
\be
p_k(x_i)=\E^{k(x_i-\mu)-\frac{\sigma^2}{2}k^2} \frac{\E^{-(x_i-\mu)^2/(2\sigma^2)}}{\sqrt{2\pi \sigma^2}}=\frac{\E^{-(x_i-\mu-\sigma^2 k)^2/(2\sigma^2)}}{\sqrt{2\pi\sigma^2}}.
\ee

\begin{figure}[t]
\resizebox{\textwidth}{!}{\includegraphics{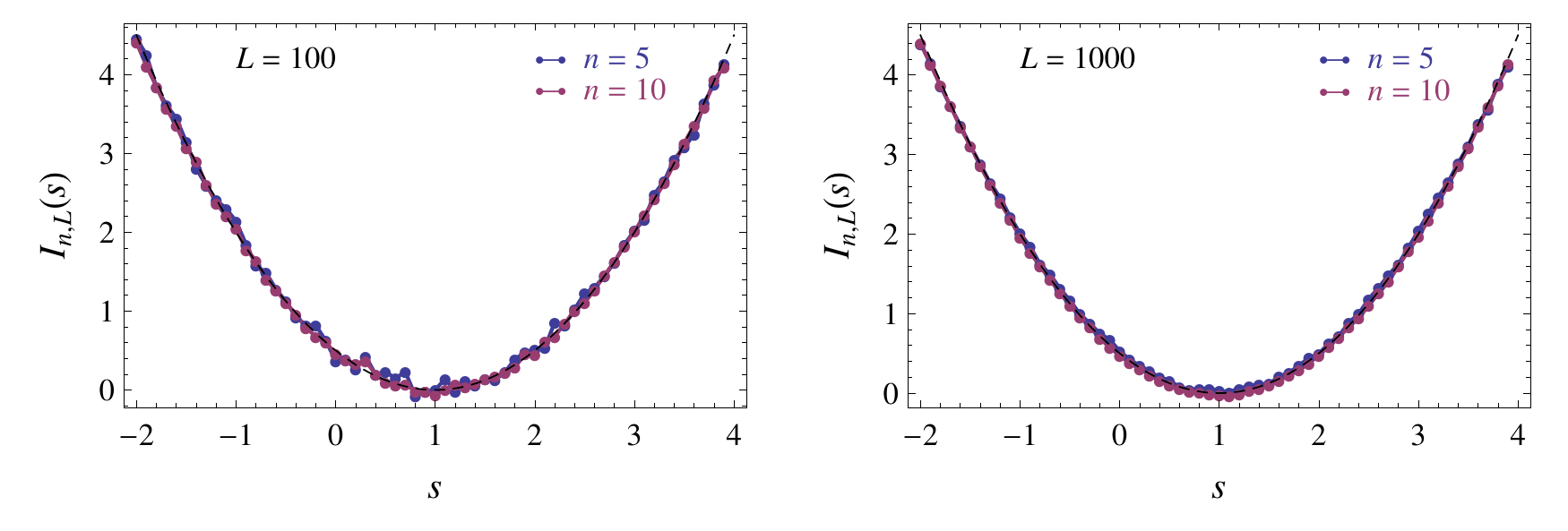}}
\caption{IS for the Gaussian IID sample mean ($\mu=\sigma=1$) with the exponential change of measure.}
\label{figexpsamp1}
\end{figure}

The estimated rate function obtained by sampling the $X_i$'s according to this Gaussian pdf with mean $\mu+\sigma^2 k$ and variance $\sigma^2$ is shown in Fig.~\ref{figexpsamp1} for various values of $n$ and $L$. This figure should be compared with Fig.~\ref{figdirsamp1} which illustrates the naive sampling method for the same sample mean. It is obvious from these two figures that the ECM method is more efficient than naive sampling.

As noted before, the rate function can be obtained in a parametric way by ``scanning'' $k$ instead of ``scanning'' $s$ and fixing $k$ according to $s$. If we were to determine $k(s)$ for a given $s$, we would find
\be
\lambda'(k)=\mu+\sigma^2 k =s,
\ee
so that $k(s)=(s-\mu)/\sigma^2$. Substituting this result in $p_k(x_i)$, then yields
\be
p_{k(s)}(x_i)=\frac{\E^{-(x_i-s)^2/(2\sigma^2)}}{\sqrt{2\pi\sigma^2}}.
\ee
Thus to efficiently sample $p_{S_n}(s)$, we must sample the $X_i$'s according to a Gaussian pdf with mean $s$ instead of the a priori pdf with mean $\mu$. The sampling is efficient in this case simply because $S_n$ concentrates in the sense of the LLN at the value $s$ instead of $\mu$, which means that $s$ has become the typical value of $S_n$ under $p_{k(s)}$. 

The idea of the ECM method is the same for processes other than IID sample means: the goal in general is to change the sampling pdf from $p$ to $p_k$ in such a way that an event that is a rare event under $p$ becomes a typical under $p_k$.\footnote{Note that this change is not always possible: for certain processes, there is no $k$ that makes certain events typical under $p_k$. For examples, see \cite{asmussen2011}.} For more information on the ECM method and IS in general; see \cite{bucklew2004} and Chaps.~V and VI of \cite{asmussen2007}. For applications of these methods, see Exercise~\ref{exis2}.

\subsection{Applications to Markov chains}

The application of the ECM method to Markov chains follows the IID case closely: we generate $L$ IID realizations $x_1^{(j)},\ldots,x_n^{(j)}$, $j=1,\ldots,L$, of the chain $\bom=X_1,\ldots, X_n$ according to the tilted joint pdf $p_k(\bom)$, using either $p_{k}$ directly or a Metropolis-type algorithm. The value $k(s)$ that achieves the efficient sampling of $S_n$ at the value $s$ is determined in the same way as in the IID case by solving $\lambda'(k)=s$, where $\lambda(k)$ is now given by Eq.~(\ref{eqscgfmc1}). In this case, $S_n=s$ becomes the typical event of $S_n$ in the limit $n\ra\infty$. A simple application of this procedure is proposed in Exercise~\ref{exmcis1}. For further reading on the sampling of Markov chains, see \cite{bucklew2004,sadowsky1989,sadowsky1990,bucklew1990b,sadowsky1996}.
 
An important difference to notice between the IID and Markov cases is that the ECM does not preserve the factorization structure of $p(\bom)$ for the latter case. In other words, $p_k(\bom)$ does not describe a Markov chain, though, as we noticed, $p_k(\bom)$ is a product pdf when $p(\bom)$ is itself a product pdf. In the case of Markov chains, $p_k(\bom)$ does retain a product structure that looks like a Markov chain, and one may be tempted to define a tilted transition pdf as
\be
\pi_k(x'|x)=\frac{\E^{kx'}\pi(x'|x)}{W(k|x)},\qquad W(k|x)=\int_{\reals} \E^{kx'} \pi(x'|x)\, \D x'.
\ee
However, it is easy to see that the joint pdf of $\bom$ obtained from this transition matrix does not reproduce the tilted pdf $p_k(\bom)$.

\subsection{Applications to continuous-time Markov processes}

The generalization of the results of the previous subsection to continuous-time Markov processes follows from our discussion of the continuous-time limit of Markov chains (see Sec.~\ref{subsecmc1}). In this case, we generate, according to the tilted pdf of the process, a sample of time-discretized trajectories. The choice of $\Delta t$ and $n$ used in the discretization will of course influence the precision of the estimates of the pdf and rate function, in addition to the sample size $L$.

Similarly to Markov chains, the tilted pdf of a continuous-time Markov process does not in general describe a new ``tilted'' Markov process having a generator related to the tilted generator $\tG_k$ defined earlier. In some cases, the tilting of a Markov process can be seen as generating a new Markov process, as will be discussed in the next subsection, but this is not true in general.

\subsection{Applications to stochastic differential equations}

Stochastic differential equations (SDEs) are covered by the continuous-time results of the previous subsection. However, for this type of stochastic processes the ECM can be expressed more explicitly in terms of the path pdf $p[x]$ introduced in Sec.~\ref{subsecpathldp}. The tilted version of this pdf, which is the functional analogue of $p_k(\bom)$, is written as
\be
p_k[x]=\frac{\E^{TkS_T[x]}\, p[x]}{W_T(k)},
\label{eqtppdf1}
\ee
where $S_T[x]$ is some functional of the trajectory $\{x(t)\}_{t=0}^T$ and 
\be
W_T(k)=E_p[\E^{TkS_T}].
\ee
The likelihood ratio associated with this change of pdf is
\be
R[x]=\frac{p[x]}{p_k[x]}=\E^{-TkS_T[x]}\, W_T(k)\approx \E^{-T[k S_T[x]-\lambda(k)]}.
\ee

As a simple application of these expressions, let us consider the SDE
\be
\dot x(t)=\xi(t),
\ee
where $\xi(t)$ is a Gaussian white noise, and the following additive process:
\be
D_T[x]=\frac{1}{T}\int_0^T \dot x(t)\, \D t,
\ee
which represents the average velocity or \emp{drift} of the SDE. What is the form of $p_k[x]$ in this case? Moreover, how can we use this tilted pdf to estimate the rate function $I(d)$ of $D_T$ in the long-time limit?

An answer to the first question is suggested by recalling the goal of the ECM. The LLN implies for the SDE above that $D_T\ra 0$ in probability as $T\ra\infty$, which means that $D_T=0$ is the typical event of the ``natural'' dynamic of $x(t)$. The tilted dynamics realized by $p_{k(d)}[x]$ should change this typical event to the fluctuation $D_T=d$; that is, the typical event of $D_T$ under $p_{k(d)}[x]$ should be $D_T=d$ rather than $D_T=0$. One candidate SDE which leads to this behavior of $D_T$ is
\be
\dot x(t)=d+\xi(t),
\label{eqbsde1}
\ee 
so the obvious guess is that $p_{k(d)}[x]$ is the path pdf of this SDE.

Let us show that this is a correct guess. The form of $p_k[x]$ according to Eq.~(\ref{eqtppdf1}) is
\be
p_k[x]\approx\E^{-T I_k[x]},\qquad I_k[x]=I[x]-k D_T[x]+\lambda(k),
\ee
where
\be
I[x]=\frac{1}{2T}\int_0^T \dot x^2\, \D t
\ee
is the action of our original SDE and
\be
\lambda(k)=\lim_{T\ra\infty}\frac{1}{T}\ln E[\E^{TkD_T}]
\ee
is the SCGF of $D_T$. The expectation entering in the definition of $\lambda(k)$ can be calculated explicitly using path integrals with the result $\lambda(k)=k^2/2$. Therefore,
\be
I_k[x]=I[x]-k D_T[x]-\frac{k^2}{2}.
\ee

From this result, we find $p_{k(d)}[x]$ by solving $\lambda'(k(d))=d$, which in this case simply yields $k(d)=d$. Hence
\be
I_{k(d)}[x]=I[x]-d D_T[x]+\frac{d^2}{2}=\frac{1}{2 T}\int_0^T (\dot x-d)^2\, \D t,
\ee
which is exactly the action of the \emp{boosted SDE} of Eq.~(\ref{eqbsde1}).

Since we have $\lambda(k)$, we can of course directly obtain the rate function $I(d)$ of $D_T$ by Legendre transform:
\be
I(d)=k(d) d-\lambda(k(d))=\frac{d^2}{2}.
\ee
This shows that the fluctuations of $D_T$ are Gaussian, which is not surprising given that $D_T$ is up to a factor $1/T$ a Brownian motion.\footnote{Brownian motion is defined as the integral of Gaussian white noise. By writing $\dot x=\xi(t)$, we therefore define $x(t)$ to be a Brownian motion.} Note that the same result can be obtained following Exercise~\ref{exisi1} by calculating the likelihood ratio $R[x]$ for a path $\{x_d(t)\}_{t=0}^T$ such that $D_T[x_d]=d$, i.e.,
\be
R[x_d]=\frac{p[x_d]}{p_{k(d)}[x_d]}\approx \E^{-T I(d)}.
\ee

These calculations give a representative illustration of the ECM method and the idea that the large deviations of an SDE can be obtained by replacing this SDE by a boosted or \emp{effective SDE} whose typical events are large deviations of the former SDE. Exercises~\ref{exaecm1} and \ref{exnecm1} show that effective SDEs need not correspond exactly to an ECM, but may in some cases reproduce such a change of measure in the large deviation limit $T\ra\infty$. The important point in all cases is to be able to calculate the likelihood ratio between a given SDE and a boosted version of it in order to obtain the desired rate function in the limit $T\ra\infty$.\footnote{The limit $\vep\ra 0$ can also be considered.}

For the particular drift process $D_T$ studied above, the likelihood ratio happens to be equivalent to a mathematical result known as \emp{Girsanov's formula}, which is often used in financial mathematics; see Exercise~\ref{exgir1}.

\subsection{Exercises}
\label{secex3}

\begin{exercise}

\item \label{exbernrv2}\exdiff{5} (Variance of Bernoulli RVs) Let $X$ be a Bernoulli RV with $P(X=1)=\alpha$ and $P(X=0)=1-\alpha$, $0\leq\alpha\leq 1$. Find the variance of $X$ in terms of $\alpha$.

\item \label{exds1}\exdiff{17} (Direct sampling of IID sample means) Generate on a computer a sample of size $L=100$ of $n$ IID Bernoulli RVs $X_1,\ldots,X_n$ with bias $\alpha=0.5$, and numerically estimate the rate function $I(s)$ associated with the sample mean $S_n$ of these RVs using the naive estimator $\hp_L(s)$ and the finite-size rate function $I_{n,L}(s)$ defined in Eq.~(\ref{eqrfe1}). Repeat for $L=10^3,10^4, 10^5$, and $10^6$ and observe how $I_{n,L}(s)$ converges to the exact rate function given in Eq.~(\ref{eqbern1}). Repeat for exponential RVs.

\item \label{exuie1}\exdiff{10} (Unbiased estimator) Prove that $\hq_L(s)$ is an unbiased estimator of $p_{S_n}(s)$, i.e., prove Eq.~(\ref{equb1}).

\item \label{exvarest1}\exdiff{15} (Estimator variance) The estimator $\hp_L(s)$ is actually a sample mean of Bernoulli RVs. Based on this, calculate the variance of this estimator with respect to $p(\bom)$ using the result of Exercise~\ref{exbernrv2} above. Then calculate the variance of the importance estimator $\hq_L(s)$ with respect to $q(\bom)$.

\item \label{exerrest1}\exdiff{15} (Relative error of estimators) The real measure of the quality of the estimator $\hp_L(s)$ is not so much its variance $\var(\hp_L(s))$ but its relative error $\err(\hp_L(s))$ defined by
\be
\err(\hp_L(s))=\frac{\sqrt{\var(\hp_L(s))}}{p_{S_n}(s)}.
\ee
Use the previous exercise to find an approximation of $\err(\hp_L(s))$ in terms of $L$ and $n$. From this result, show that the sample size $L$ needed to achieve a certain relative error grows exponentially with $n$. (Source: Sec.~V.I of \cite{asmussen2007}.)

\item \label{exoptq1}\exdiff{17} (Optimal importance pdf) Show that the density $q^*$ defined in (\ref{eqoippdf1}) is optimal in the sense that
\be
\var_q(\hq_L(s))\geq\var_{q^*}(\hq_L(s))=0
\ee
for all $q$ relatively continuous to $p$. (See Chap.~V of \cite{asmussen2007} for help.)

\item \label{exisscgf1}\exdiff{15} (Exponential change of measure) Show that the value of $k$ solving Eq.~(\ref{eqecm1}) is given by $\lambda'(k)=s$, where $\lambda(k)$ is the SCGF defined in Eq.~(\ref{eqscgfis1}).

\item \exdiff{40} (LDP for estimators) We noticed before that $\hp_L(s)$ is the empirical vector of the IID sample $\{S_n^{(j)}\}_{j=1}^L$. Based on this, we could study the LDP of $\hp_L(s)$ rather than just its mean and variance, as usually done in sampling theory. What is the form of the LDP associated with $\hp_L(s)$ with respect to $p(\bom)$? What is its rate function? What is the minimum and zero of that rate function? Answer the same questions for $\hq_L(s)$ with respect to $q(\bom)$. (Hint: Use Sanov's Theorem and the results of Exercise~\ref{exild1}.)

\item \label{exis2}\exdiff{25} (Importance sampling of IID sample means) Implement the ECM method to numerically estimate the rate function associated with the IID sample means of Exercise~\ref{exiid1}. Use a simple IID sampling based on the explicit expression of $p_k$ in each case (i.e., assume that $W(k)$ is known). Study the convergence of the results as a function of $n$ and $L$ as in Fig.~\ref{figexpsamp1}.

\item \label{exisi1}\exdiff{15} (IS estimator) Show that $\hq_L(s)$ for $p_{k(s)}$ has the form
\be
\hq_L(s)\approx\frac{1}{L\Delta s}\sum_{j=1}^L \id_{\Delta_s}(s^{(j)})\ \E^{-nI(s^{(j)})}.
\ee
Why is this estimator trivial for estimating $I(s)$?

\item \label{exmetis1}\exdiff{30} (Metropolis sampling of IID sample means) Repeat Exercise~\ref{exis2}, but instead of using an IID sampling method, use the Metropolis algorithm to sample the $X_i$'s in $S_n$. (For information on the Metropolis algorithm, see Appendix~\ref{appmh}.)

\item \exdiff{15} (Tilted distribution from contraction) Let $X_1,\ldots,X_n$ be a sequence of real IID RVs with common pdf $p(x)$, and denote by $S_n$ and $L_n(x)$ the sample mean and empirical functional, respectively, of these RVs. Show that the ``value'' of $L_n(x)$ solving the minimization problem associated with the contraction of the LDP of $L_n(x)$ down to the LDP of $S_n$ is given by
\be
\mu_k(x)=\frac{\E^{kx}p(x)}{W(k)},\qquad W(k)=E_p[\E^{kX}],
\ee
where $k$ is such that $\lambda'(k)=(\ln W(k))'=s$. To be more precise, show that this pdf is the solution of the minimization problem
\be
\inf_{\mu:f(\mu)=s} I(\mu),
\ee
where
\be
I(\mu)= \int_\reals\D x\, \mu(x)\ln\frac{\mu(x)}{p(x)}
\ee
is the continuous analogue of the relative entropy defined in Eq.~(\ref{eqsanov2}) and
\be
f(\mu)=\int_\reals x\, \mu(x)\, \D x
\ee
is the contraction function. Assume that $W(k)$ exists. Why is $\mu_k$ the same as the tilted pdf used in IS?

\item \exdiff{50} (Equivalence of ensembles) Comment on the meaning of the following statements: (i) The optimal pdf $q^*$ is to the microcanonical ensemble what the tilted pdf $p_k$ is to the canonical ensemble. (ii) Proving that $p_k$ achieves a zero variance for the estimator $\hq_L(s)$ in the limit $n\ra\infty$ is the same as proving that the canonical ensemble becomes equivalent to the microcanonical ensemble in the thermodynamic limit.

\item \label{exmcis1}\exdiff{25} (Tilted Bernoulli Markov chain) Find the expression of the tilted joint pdf $p_k(\bom)$ for the Bernoulli Markov chain of Exercise~\ref{exbmc1}. Then use the Metropolis algorithm (see Appendix~\ref{appmh}) to generate a large-enough sample of realizations of $\bom$ in order to estimate $p_{S_n}(s)$ and $I(s)$. Study the converge of the estimation of $I(s)$ in terms of $n$ and $L$ towards the expression of $I(s)$ found in Exercise~\ref{exbmc1}. Note that for a starting configuration $x_1,\ldots,x_n$ and a target configuration $x'_1,\ldots,x'_n$, the acceptance ratio is
\be
\frac{p_k(x_1',\ldots,x_n')}{p_k(x_1,\ldots,x_n)}=\frac{\E^{k x_1'} p(x_1')\prod_{i=2}^n \E^{k x_i'} \pi(x_i'|x_{i-1}')}{\E^{k x_1} p(x_1)\prod_{i=2}^n \E^{k x_i} \pi(x_i|x_{i-1})},
\ee
where $p(x_i)$ is some initial pdf for the first state of the Markov chain. What is the form of $\hq_L(s)$ in this case?

\item \label{exgir1}\exdiff{20} (Girsanov's formula) Denote by $p[x]$ the path pdf associated with the SDE 
\be
\dot x(t)=\sqrt{\vep}\, \xi(t),
\ee
where $\xi(t)$ is Gaussian white noise with unit noise power. Moreover, let $q[x]$ be the path pdf of the boosted SDE
\be
\dot x(t)=\mu +\sqrt{\vep}\, \xi(t).
\ee
Show that
\be
R[x]=\frac{p[x]}{q[x]}=\exp\left(-\frac{\mu}{\vep} \int_0^T \dot x\, \D t+\frac{\mu^2}{2\vep} T\right)=\exp\left(-\frac{\mu}{\vep} x(T) +\frac{\mu^2}{2\vep} T\right).
\ee

\item \label{exaecm1}\exdiff{27} (Effective SDE) Consider the additive process,
\be
S_T=\frac{1}{T}\int_0^T x(t)\, \D t,
\ee
with $x(t)$ evolving according to the simple Langevin equation of Exercise~\ref{exoup1}. Show that the tilted path pdf $p_{k(d)}[x]$ associated with this process is \textit{asymptotically} equal to the path pdf of the following boosted SDE:
\be
\dot x(t)=-\gamma (x(t)-s)+\xi(t).
\label{eqbsde2}
\ee 
More precisely, show that the action $I_{k(s)}[x]$ of $p_{k(s)}[x]$ and the action $J[x]$ of the boosted SDE differ by a boundary term which is of order $O(1/T)$.

\item \label{exnecm1}\exdiff{22} (Non-exponential change of measure) Although the path pdf of the boosted SDE of Eq.~(\ref{eqbsde2}) is not exactly the tilted path pdf $p_{k}[x]$ obtained from the ECM, the likelihood ratio $R[x]$ of these two pdfs does yield the rate function of $S_T$. Verify this by obtaining $R[x]$ exactly and by using the result of Exercise~\ref{exisi1}.

\end{exercise} 

\section{Other numerical methods for large deviations}
\label{secnum2}

The use of the ECM method appears to be limited, as mentioned before, by the fact that the tilted pdf $p_k(\bom)$ involves $W_n(k)$. We show in this section how to circumvent this problem by considering estimators of $W_n(k)$ and $\lambda(k)$ instead of estimators of the pdf $p({S_n})$, and by sampling these new estimators according to the tilted pdf $p_k(\bom)$, but with the Metropolis algorithm in order not to rely on $W_n(k)$, or directly according to the a priori pdf $p(\bom)$. At the end of the section, we also briefly describe other important methods used in numerical simulations of large deviations. 

\subsection{Sample mean method}

We noted in the previous subsection that the parameter $k$ entering in the ECM had to be chosen by solving $\lambda'(k)=s$ in order for $s$ to be the typical event of $S_n$ under $p_k$. The reason for this is that
\be
\lim_{n\ra\infty }E_{p_k}[S_n]=\lambda'(k).
\ee
By the LLN we also have that $S_n\ra\lambda'(k)$ in probability as $n\ra\infty$ if $S_n$ is sampled according to $p_{k}$.

These results suggest using
\be
s_{L,n}(k)=\frac{1}{L}\sum_{j=1}^L S_n(\bom^{(j)})
\ee
as an estimator of $\lambda'(k)$ by sampling $\bom$ according to $p_k(\bom)$, and then integrating this estimator with the boundary condition $\lambda(0)=0$ to obtain an estimate of $\lambda(k)$. In other words, we can take our estimator for $\lambda(k)$ to be
\be
\hlambda_{L,n}(k)=\int_0^k s_{L,n}(k')\, \D k'.
\ee
From this estimator, we then obtain a parametric estimation of $I(s)$ from the GE Theorem by Legendre-transforming $\hlambda_{L,n}(k)$:
\be
I_{L,n}(s)=k s-\hlambda_{L,n}(k),
\label{eqplf1}
\ee
where $s=s_{L,n}(k)$ or, alternatively,
\be
I_{L,n}(s)=k(s)\, s-\hlambda_{L,n}(k(s)),
\label{eqplf2}
\ee
where $k(s)$ is the root of $\hlambda_{L,n}'(k)=s$.\footnote{In practice, we have to check numerically (using finite-scale analysis) that $\lambda_{n,L}(k)$ does not develop non-differentiable points in $k$ as $n$ and $L$ are increased to $\infty$. If non-differentiable points arise, then $I(s)$ is recovered from the GE Theorem only in the range of $\lambda'$; see Sec.~4.4 of \cite{touchette2009}.}

The implementation of these estimators with the Metropolis algorithm is done explicitly via the following steps: 
\begin{enumerate}
\item For a given $k\in\reals$, generate an IID sample $\{\bom^{(j)}\}_{j=1}^L$ of $L$ configurations $\bom=x_1,\ldots,x_n$ distributed according to $p_{k}(\bom)$ using the Metropolis algorithm, which only requires the computation of the ratio
\be
\frac{p_{k}(\bom)}{p_{k}(\bom')}=\frac{\E^{nkS_n(\bom)}\, p(\bom)}{\E^{nkS_n(\bom')}\, p(\bom')}
\ee
for any two configurations $\bom$ and $\bom'$; see Appendix~\ref{appmh};

\item Compute the estimator $s_{L,n}(k)$ of $\lambda'(k)$ for the generated sample;
\item Repeat the two previous steps for other values of $k$ to obtain a numerical approximation of the function $\lambda'(k)$;
\item Numerically integrate the approximation of $\lambda'(k)$ over the mesh of $k$ values starting from $k=0$. The result of the integration is the estimator $\hlambda_{L,n}(k)$ of $\lambda(k)$;
\item Numerically compute the Legendre transform, Eq.~(\ref{eqplf1}), of the estimator of $\lambda(k)$ to obtain the estimator $I_{L,n}(s)$ of $I(s)$ at $s=s_{L,n}(k)$;
\item Repeat the previous steps for larger values of $n$ and $L$ until the results converge to some desired level of accuracy.
\end{enumerate}

\begin{figure}[t]
\centering
\resizebox{\textwidth}{!}{\includegraphics{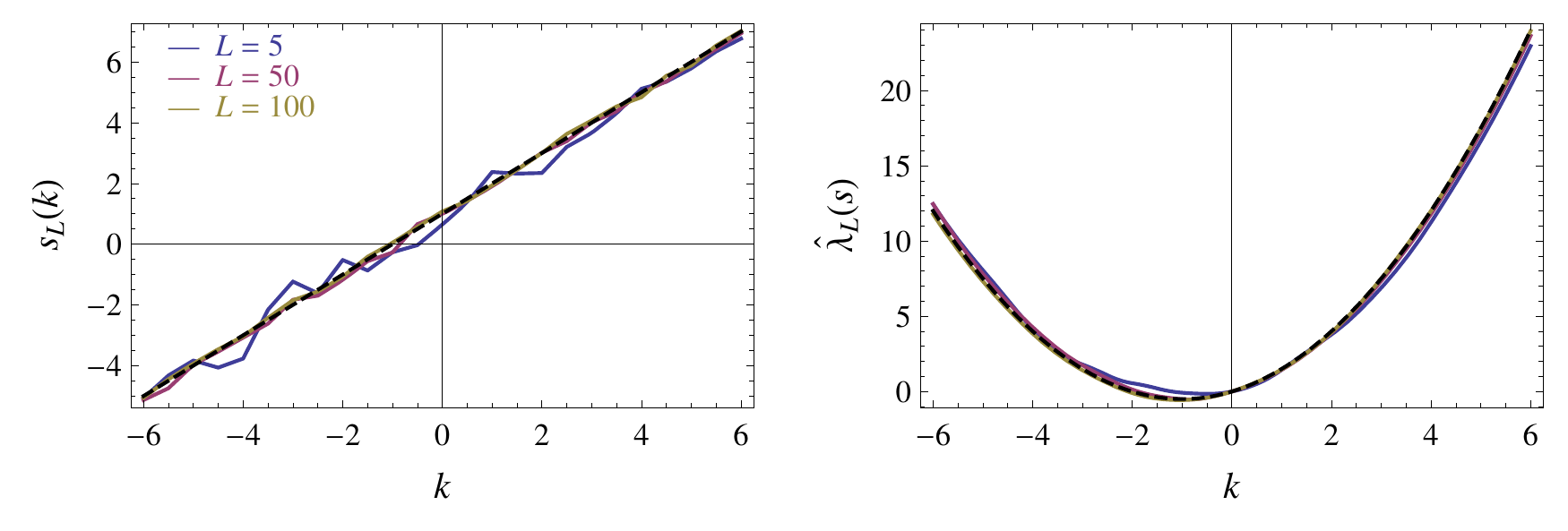}}
\resizebox{0.5\textwidth}{!}{\includegraphics{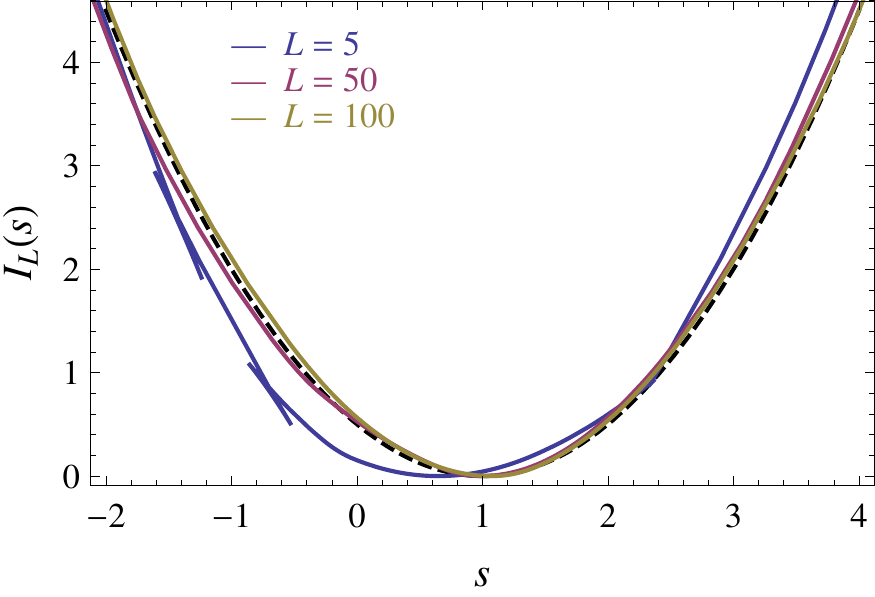}}
\caption{(Top left) Empirical sample mean $s_L(k)$ for the Gaussian IID sample mean ($\mu=\sigma=1$). The spacing of the $k$ values is $\Delta k=0.5$. (Top right) SCGF $\hlambda_L(k)$ obtained by integrating $s_L(k)$. (Bottom) Estimate of $I(s)$ obtained from the Legendre transform of $\hlambda_L(k)$. The dashed line in each figure is the exact result. Note that the estimate of the rate function for $L=5$ is multi-valued because the corresponding $\hlambda_L(k)$ is nonconvex.}
\label{figgaussiansm1}
\end{figure}

The result of these steps are illustrated in Fig.~\ref{figgaussiansm1} for our test case of the IID Gaussian sample mean for which $\lambda'(k)=\mu+\sigma^2 k$. Exercise~\ref{exsmm1} covers other sample means studied before (e.g., exponential, binary, etc.).

Note that for IID sample means, one does not need to generate $L$ realizations of the whole sequence $\bom$, but only $L$ realizations of one summand $X_i$ according the tilted marginal pdf $p_k(x)=\E^{kx} p(x)/W(k)$. In this case, $L$ takes the role of the large deviation parameter $n$. For Markov chains, realizations of the whole sequence $\bom$ must be generated one after the other, e.g., using the Metropolis algorithm.

\subsection{Empirical generating functions}

The last method that we cover is based on the estimation of the generating function
\be
W_n(k)=E[\E^{nk S_n}]=\int_{\reals^n}p(\bom)\, \E^{nkS_n(\bom)}\, \D\bom.
\ee
Consider first the case where $S_n$ is an IID sample mean. Then $\lambda(k)=\ln W(k)$, where $W(k)=E[\E^{kX_i}]$, as we know from Sec.~\ref{secsanov}, and so the problem reduces to the estimation of the SCGF of the common pdf $p(X)$ of the $X_i$'s. An obvious estimator for this function is
\be
\hlambda_{L}(k)=\ln \frac{1}{L}\sum_{j=1}^L \E^{k X^{(j)}},
\ee 
where $X^{(j)}$ are the IID samples drawn from $p(X)$. From this estimator, which we call the \emp{empirical SCGF}, we build a parametric estimator of the rate function $I(s)$ of $S_n$ using the Legendre transform of Eq.~(\ref{eqplf1}) with
\be
s(k)=\hlambda_L'(k)=\frac{\sum_{j=1}^L X^{(j)}\, \E^{k X^{(j)}}}{\sum_{j=1}^L \E^{k X^{(j)}}}.
\ee

\begin{figure}[t]
\centering
\resizebox{\textwidth}{!}{\includegraphics{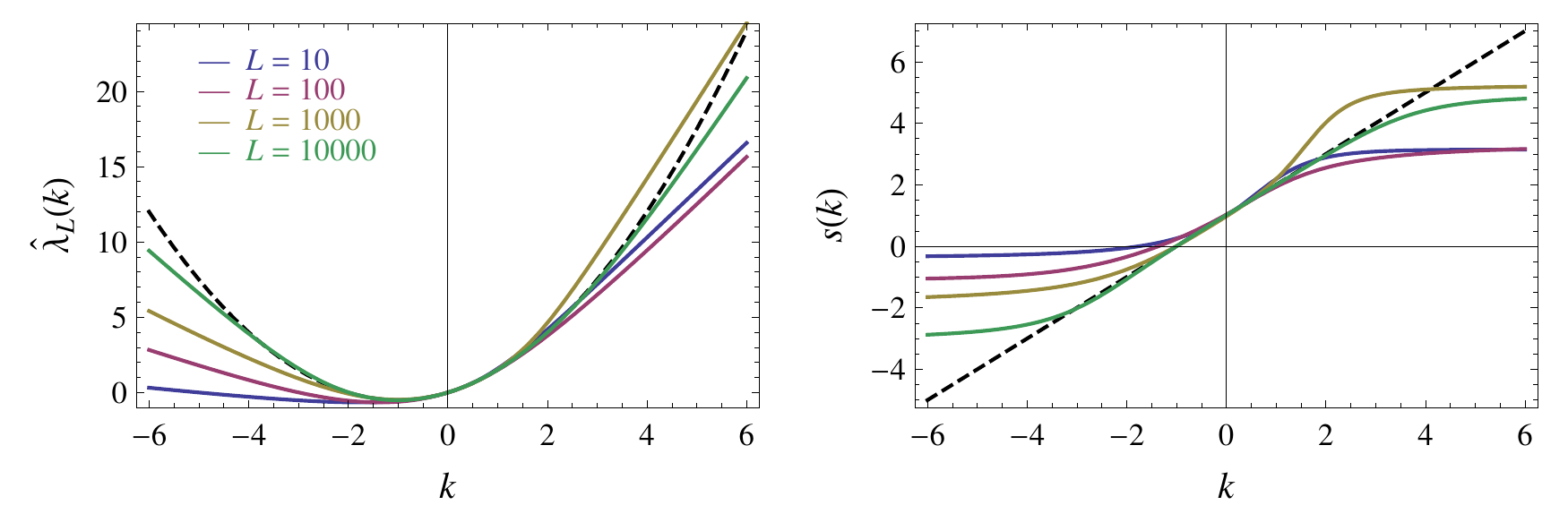}}
\resizebox{0.5\textwidth}{!}{\includegraphics{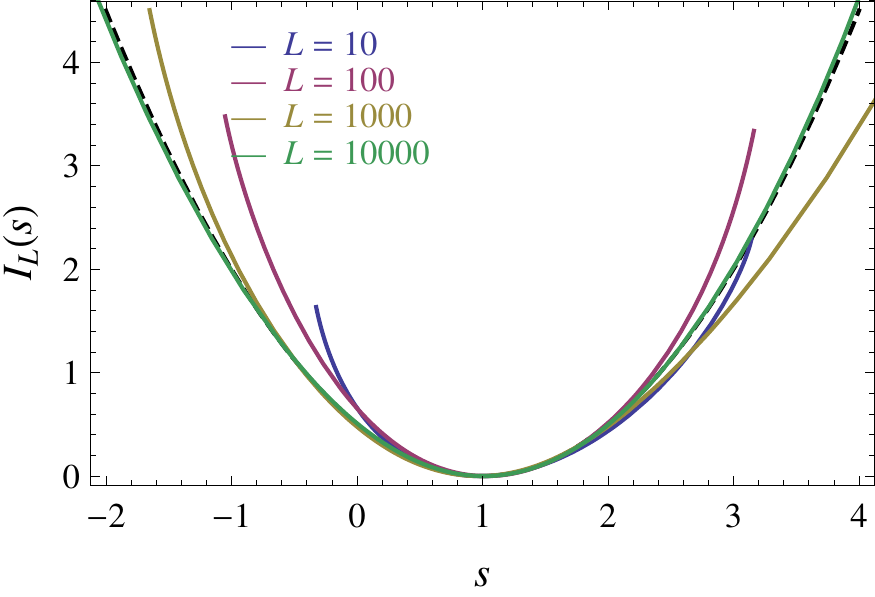}}
\caption{(Top left) Empirical SCGF $\hlambda_L(k)$ for the Gaussian IID sample mean ($\mu=\sigma=1$). (Top right)  $s(k)=\hlambda_L'(k)$. (Bottom) Estimate of $I(s)$ obtained from $\hlambda_L(k)$. The dashed line in each figure is the exact result.}
\label{figgaussianegf1}
\end{figure}

Fig.~\ref{figgaussianegf1} shows the result of these estimators for the IID Gaussian sample mean. As can be seen, the convergence of $I_L(s)$ is fast in this case, but is limited to the central part of the rate function. This illustrates one limitation of the empirical SCGF method: for unbounded RVs, such as the Gaussian sample mean, $W(k)$ is correctly recovered for large $|k|$ only for large samples, i.e., large $L$, which implies that the tails of $I(s)$ are also recovered only for large $L$. For bounded RVs, such as the Bernoulli sample mean, this limitation does not arise; see \cite{duffield1995,duffy2005} and Exercise~\ref{exegf1}.

The application of the empirical SCGF method to sample means of ergodic Markov chains is relatively direct. In this case, although $W_n$ no longer factorizes into $W(k)^n$, it is possible to ``group'' the $X_i$'s into blocks of $b$ RVs as follows:
\be
\underbrace{X_1+\cdots +X_b}_{Y_1} + \underbrace{X_{b+1}+\cdots+X_{2b}}_{Y_2}+\cdots+\underbrace{X_{n-b+1}+\cdots+X_n}_{Y_m}
\ee
where $m=n/b$, so as to rewrite the sample mean $S_n$ as
\be
S_n=\frac{1}{n}\sum_{i=1}^n X_i=\frac{1}{bm}\sum_{i=1}^m Y_i.
\ee
If $b$ is large enough, then the blocks $Y_i$ can be treated as being independent. Moreover, if the Markov chain is ergodic, then the $Y_i$'s should be identically distributed for $i$ large enough. As a result, $W_n(k)$ can be approximated for large $n$ and large $b$ (but $b\ll n$) as
\be
W_n(k)=E[\E^{k\sum_{i=1}^m Y_i}]\approx E[\E^{k Y_i}]^m,
\ee  
so that
\be
\lambda_n(k)=\frac{1}{n}\ln W_n(k)\approx \frac{m}{n}\ln E[\E^{k Y_i}]=\frac{1}{b}\ln E[\E^{k Y_i}].
\ee
We are thus back to our original IID problem of estimating a generating function; the only difference is that we must perform the estimation at the level of the $Y_i$'s instead of the $X_i$'s. This means that our estimator for $\lambda(k)$ is now
\be
\hlambda_{L,n}(k)=\frac{1}{b}\ln \frac{1}{L}\sum_{j=1}^L \E^{k Y^{(j)}},
\ee
where $Y^{(j)}$, $j=1,\ldots, L$ are IID samples of the block RVs. Following the IID case, we then obtain an estimate of $I(s)$ in the usual way using the Legendre transform of Eq.~(\ref{eqplf1}) or (\ref{eqplf2}).

The application of these results to Markov chains and SDEs is covered in Exercises~\ref{exegf1} and \ref{exegf2}. For more information on the empirical SCGF method, including a more detailed analysis of its convergence, see \cite{duffield1995,duffy2005}.

\subsection{Other methods}

We briefly describe below a number of other important methods used for estimating large deviation probabilities, and give for each of them some pointer references for interested readers. The fundamental ideas and concepts related to sampling and ECM that we have treated before also arise in these other methods.
\begin{itemize}
\item \emp{Splitting methods:} Splitting or cloning methods are the most used sampling methods after Monte Carlo algorithms based on the ECM. The idea of splitting is to ``grow'' a sample or ``population'' $\{\bom\}$ of sequences or configurations in such a way that a given configuration $\bom$ appears in the population in proportion to its weight $p_k(\bom)$. This is done in some iterative fashion by ``cloning'' and ``killing'' some configurations. The result of this process is essentially an estimate for $W_n(k)$, which is then used to obtain $I(s)$ by Legendre transform. For mathematically-oriented introductions to splitting methods, see \cite{lecuyer2007,lecuyer2009b,dean2009,morio2010}; for more physical introductions, see \cite{giardina2006,lecomte2007a,tailleur2007,tailleur2009}.

\item \emp{Optimal path method:} We briefly mentioned in Sec.~\ref{subsecpathldp} that rate functions related to SDEs can often be derived by contraction of the action functional $I[x]$. The optimization problem that results from this contraction is often not solvable analytically, but there are several techniques stemming from optimization theory, classical mechanics, and dynamic programming that can be used to solve it. A typical optimization problem in this context is to find the optimal path that a noisy system follows with highest probability to reach a certain state considered to be a rare event or fluctuation. This optimal path is also called a \emp{transition path} or a \emp{reactive trajectory}, and can be found analytically only in certain special systems, such as linear SDEs and conservative systems; see Exercises~\ref{exop1}--\ref{exop3}. For nonlinear systems, the Lagrangian or Hamiltonian equations governing the dynamics of these paths must be simulated directly; see \cite{graham1989} and Sec.~6.1 of \cite{touchette2009} for more details.

\item \emp{Transition path method:} The transition path method is essentially a Monte Carlo sampling method in the space of configurations $\bom$ (for discrete-time processes) or trajectories $\{x(t)\}_{t=0}^T$ (for continuous-time processes and SDEs), which aims at finding, as the name suggests, transition paths or optimal paths. The contribution of Dellago in this volume covers this method in detail; see also \cite{dellago2003,dellago2006,dellago2009,metzner2006,vanden-eijnden2006b}. A variant of the transition method is the so-called \emp{string method} \cite{vanden-eijnden2002} which evolves paths in conservative systems towards optimal paths connecting different local minima of the potential or landscape function of these systems.

\item \emp{Eigenvalue method:} We have noted that the SCGF of an ergodic continuous-time process is given by the dominant eigenvalue of its tilted generator. From this connection, one can attempt to numerically evaluate SCGFs using numerical approximation methods for linear functional operators, combined with numerical algorithms for finding eigenvalues. A recent application of this approach, based on the renormalization group, can be found in \cite{gorissen2009,gorissen2011}.
\end{itemize}

\subsection{Exercises}
\label{secex4}

\begin{exercise}
\item \label{exsmm1}\exdiff{25} (Sample mean method) Implement the steps of the sample mean method to numerically estimate the rate functions of all the IID sample means studied in these notes, including the sample mean of the Bernoulli Markov chain.

\item \exdiff{25}\label{exegf1} (Empirical SCGF method) Repeat the previous exercise using the empirical SCGF method instead of the sample mean method. Explain why the former method converges quickly when $S_n$ is bounded, e.g., in the Bernoulli case. Analyze in detail how the method converges when $S_n$ is unbounded, e.g., for the exponential case. (Source: \cite{duffield1995,duffy2005}.)

\item \exdiff{30}\label{exegf2} (Sampling of SDEs) Generalize the sample mean and empirical mean methods to SDEs. Study the additive process of Exercise~\ref{exaecm1} as a test case. 

\item \exdiff{30}\label{exop1} (Optimal paths for conservative systems) Consider the conservative system of Exercise~\ref{excons1} and assume, without loss of generality, that the global minimum of the potential $U(\bx)$ is at $\bx=0$. Show that the optimal path which brings the system from $\bx(0)=0$ to some fluctuation $\bx(T)=\bx\neq 0$ after a time $T$ is the time-reversal of the solution of the noiseless dynamics $\dot \bx =-\nabla U(\bx)$ under the terminal conditions $\bx(0)=\bx$ and $\bx(T)=0$. The latter path is often called the \emp{decay path}. 

\item \exdiff{25}\label{exop2} (Optimal path for additive processes) Consider the additive process $S_T$ of Exercise~\ref{exaecm1} under the Langevin dynamics of Exercise~\ref{exoup1}. Show that the optimal path leading to the fluctuation $S_T=s$ is the constant path
\be
x(t)=s,\quad t\in [0,T].
\ee
Explain why this result is the same as the asymptotic solution of the boosted SDE of Eq.~(\ref{eqbsde2}).

\item \exdiff{25}\label{exop3} (Optimal path for the dragged Brownian particle) Apply the previous exercise to Exercise~\ref{exbmp1}. Discuss the physical interpretation of the results.

\end{exercise} 

\section*{Acknowledgments}

I would like to thank Rosemary J. Harris for suggesting many improvements to these notes, as well as Alexander Hartmann and Reinhard Leidl for the invitation to participate to the 2011 Oldenburg Summer School.

\appendix
\section{Metropolis-Hastings algorithm}
\label{appmh}

The \emp{Metropolis-Hastings algorithm}\footnote{Or, more precisely, the Metropolis--Rosenbluth--Rosenbluth--Teller--Teller--Hastings algorithm.} or \emp{Markov chain Monte Carlo algorithm} is a simple method used to generate a sequence $x_1,\dots,x_n$ of variates whose empirical distribution $L_n(x)$ converges asymptotically to a \emp{target pdf} $p(x)$ as $n\ra\infty$. The variates are generated sequentially in the manner of a Markov chain as follows:
\begin{enumerate}
\item Given the value $x_i=x$, choose another value $x'$, called the \emp{proposal} or \emp{move}, according to some (fixed) conditional pdf $q(x'|x)$, called the \emp{proposal pdf};
\item Accept the move $x'$, i.e., set $x_{i+1}=x'$ with probability $\min\{1,a\}$ where
\be
a=\frac{p(x')\, q(x|x')}{p(x)\, q(x'|x)};
\ee
\item If the move is not accepted, set $x_{i+1}=x$.
\end{enumerate}

This algorithm is very practical for sampling pdfs having an exponential form, such as the Boltzmann-Gibbs distribution or the tilted pdf $p_k$, because it requires only the computation of the \emp{acceptance ratio} $a$ for two values $x$ and $x'$, and so does not require the knowledge of the normalization constant that usually enters in these pdfs. For the tilted pdf $p_k(\bom)$, for example, 
\be
a=\frac{p_k(\bom')\, q(\bom|\bom')}{p_k(\bom)\, q(\bom'|\bom)}=\frac{\E^{nk S_n(\bom)}\, p(\bom)\, q(\bom|\bom')}{\E^{nkS_n(\bom)}\, p(\bom)\, q(\bom'|\bom)}.
\ee

The original \emp{Metropolis algorithm} commonly used in statistical physics is obtained by choosing $q(x'|x)$ to be symmetric, i.e., if $q(x'|x)=q(x|x')$ for all $x$ and $x'$, in which case $a=p(x')/p(x)$. If $q(x'|x)$ does not depend on $x$, the algorithm is referred to as the \emp{independent chain Metropolis-Hastings algorithm}.

For background information on Monte Carlo algorithms, applications, and technical issues such as mixing, correlation and convergence, see \cite{krauth2006} and Chap.~13 of \cite{asmussen2007}.







\newpage
\section*{Errata}

I list below some errors contained in the published version of these notes and in the versions (v1 and v2) previously posted on the arxiv.

The numbering of the equations in the published version is different from the version posted on the arxiv. The corrections listed here refer to the arxiv version.

I thank Bernhard Altaner and Andreas Engel for reporting errors. I also thank Vivien Lecomte for a discussion that led me to spot the errors of Sections 4.4 and 4.5.

\subsection*{Corrections to arxiv version 1}

\begin{itemize}
\item p.~15, in Eq.~(2.35): The correct result is $\alpha^{sn}(1-\alpha)^{(1-s)n}$. [Noted by B. Altaner]
\item p.~16, in Ex.~2.7.8: ``$\lambda(k)$ has a continuous derivative'' added.
\item p.~20, in Eq.~(3.15): The limit $T\rightarrow 0$ should be $T\rightarrow\infty$.  [Noted by B. Altaner]

\item p.~25, in Ex.~3.6.2: $Y\in \{-1,+1\}$ with $P(Y=1)=P(Y=-1)=1/2$ instead of Bernoulli.

\item p.~25, in Ex.~3.6.5: The RVs are now specified to be Gaussian RVs.
\end{itemize}

\subsection*{Corrections to arxiv version 2}

\begin{itemize}
\item p.~1: In the abstract: missing `s' to `introduce'.

\item p.~16, in Ex.~2.7.8: The conditions of $\lambda(k)$ stated in the exercise are again wrong: $\lambda(k)$ must be differentiable and must not have affine parts to ensure that $\lambda'(k)=s$ has a unique solution.

\item p.~20, in Eq.~(3.17): The tilted generator is missing an identity matrix; it 
should read
\[
\tilde G_k=G+k x' \delta_{x,x'}.
\]
[Noted by A. Engel] Part of Section 3.3 and Eq.~(3.17) was rewritten to emphasize where this identity matrix term is coming from.

\item p.~27, in Eq.~(3.40): The generator should have the off diagonal elements exchanged. Idem for the matrix shown in Eq.~(3.45). [Noted by A. Engel] 

These notes represent probability vectors as columns vectors. Therefore, the columns, not the rows, of stochastic matrices must sum to one, which means that the columns, not the rows, of stochastic generators must sum to zero.

\item p.~27, in Ex.~3.6.13: It should be mentioned that $f(x_i,x_{i+1})$ is chosen as $f(x_i,x_{i+1})=1-\delta_{x_i,x_{i+1}}$. [Noted by A. Engel]

\item p.~36, in Section 4.4: Although the tilted pdf $p_k(\bom)$ of a Markov chain has a product structure reminiscent of a Markov chain, it cannot be described as a Markov chain, contrary to what is written. In particular, the matrix shown in Eq.~(4.24) is not a proper stochastic matrix with columns summing to one. For that, we should define the matrix
\[
\pi_k(x'|x)=\frac{e^{kx'}\pi(x'|x)}{W(k|x)},\qquad W(k|x)=\int e^{kx'}\pi(x'|x)\, dx'.
\]
However, the pdf of $\bom$ constructed from this transition matrix is different from the tilted joint pdf $p_k(\bom)$. Section 4.4 was rewritten to correct this error.

This section is not entirely wrong: the Metropolis sampling can still be done at the level of $p_k$ by generating IID sequences $\bom^{(j)}$, as now described.

\item p.~37, in Section 4.5: The error of Section 4.4 is repeated here at the level of generators: the $G_k$ shown in Eq.~(4.25) is not a proper stochastic generator with ``columns'' summing to zero. Section 4.5 was completely rewritten to correct this error.

\item p.~42, Ex.~4.7.14: This exercise refers to the tilted joint pdf $p_k(\bom)$; there is no tilted transition matrix here. The exercise was merged with the next one to correct this error.
\end{itemize}

\end{document}